\documentclass[usenatbib]{mnras}
\usepackage{graphicx}
\usepackage{amssymb, amsmath}
\usepackage{times}
\usepackage{color}
\usepackage{multirow}
\usepackage{lscape}
\usepackage{varwidth}
\usepackage[section]{placeins}
\usepackage{epstopdf}
\usepackage{epsf}
\usepackage{float}
\usepackage[normalem]{ulem}
\usepackage{booktabs}
\usepackage{hyperref}
\usepackage{academicons}
\usepackage{xcolor}

\newcommand{\cmnt}[1]{}

\newcommand{\HI}{\ion{H}{i}}
\newcommand{\lya}{Lyman-$\alpha$}

\begin{document}
\setlength{\parskip}{0pt}

\title[\lya\ $\times$ Galaxy on large scale UVB fluctuation]{Probing Large Scale Ionizing Background Fluctuation with Lyman $\alpha$ Forest and Galaxy Cross-correlation at z=2.4}
\author[Long \& Hirata]{Heyang Long$^{1,2}$\thanks{E-mail: long.1697@osu.edu} and
Christopher M. Hirata$^{1,2,3}$\thanks{E-mail: hirata.10@osu.edu}\\
$^1$Department of Physics, The Ohio State University, 191 West Woodruff Avenue, Columbus, Ohio 43210, USA\\
$^2$Center for Cosmology and AstroParticle Physics (CCAPP), The Ohio State University, 191 West Woodruff Avenue, Columbus, Ohio 43210, USA\\
$^3$Department of Astronomy, The Ohio State University, 140 West 18th Avenue, Columbus, Ohio 43210, USA
}

\maketitle

\begin{abstract}
The amplitude of the metagalactic ultraviolet background (UVB) at large-scales is impacted by two factors. First, it naturally attenuates at scales larger than mean-free-path of UVB photons due to the absorption by neutral intergalactic medium. Second, there are discrete and rare ionizing sources distributing in the Universe, emitting the UVB photons, and thus enhancing the local UVB amplitude. Therefore, for cosmological probe that is sensitive to the UVB amplitude and capable of detecting the large scale like \lya\ forest spectrum, the fluctuation due to the clustering of ionizing sources becomes a significant factor for \lya\ flux transmission and leave imprints on \lya\ flux power spectrum at these large scales. In this work, we make use of a radiative transfer model that parametrizes the UVB source distribution by its bias $b_{\rm j}$ and shot noise $\overline{n}_{\rm j}$. We estimate the constraints on this model through the cross-correlation between \lya\ forest survey and galaxy survey, using the DESI \lya\ forest survey and the Roman Space Telescope emission line galaxy survey as an example. We show the detection sensitivity improvement for UVB parameters from disjoint to maximal overlap of DESI+Roman survey strategy. We also show that the degeneracy of two ionizing source parameters can be broken by increasing the overlapping survey area. Our results motivate survey strategies more dedicated to probe the UVB large-scale fluctuations. 
\end{abstract}

\begin{keywords}
intergalactic medium - cosmology: large-scale structure of Universe
\end{keywords}

\section{Introduction}

The intergalactic medium (IGM) after the Epoch of Reionization (EoR) is maintained highly ionized by metagalactic ultraviolet background (UVB), which is generated by star-forming galaxies (SFGs) and quasars (see \citealt{2022ARA&A..60..121R,2019arXiv190706653W,2019MNRAS.485...47P, 2016ARA&A..54..313M} for a review). Determining the amplitude and spectral shape of UVB as a function of redshift is critical to understand relative contributions from ionizing sources \citep{2008ApJ...688...85F}. Moreover, in practice the post-reionization thermal and ionization state of intergalactic medium (IGM) and structure formation is usually simulated in light of a spatially uniform UVB.

There are various ways to measure UVB through cosmic epoch. The proximity effect has been a classical probe of UVB intensity for years \citep{1988ApJ...327..570B,1987ApJ...319..709C,2010ApJ...722..699D,2008A&A...491..465D,2011MNRAS.412.2543C}. Luminous quasars could produce large amounts of ionizing photons and thus enhance the transmission of the Lyman-$\alpha$ forest near themselves (up to several Mpc) \citep{2011MNRAS.410.1096B}. While quasar luminosity determines the enhanced portion of transmission, the size of this ``proximity zone'' also depends on UVB intensity (e.g. \citealt{ 2008A&A...491..465D, 2011MNRAS.412.2543C,Chen:2020uhh}). So with quasar luminosity known, the UVB intensity could be inversely deduced by comparing \lya\ spectra in the proximity zone to those in the absence of quasar at the same redshift.
An alternative way of estimating UVB intensity is developed by comparing simulated \lya\ forest spectra to the real observation data \citep{2005pgqa.conf..219B,2007MNRAS.382..325B,2013MNRAS.436.1023B,2019MNRAS.486..769K,2021MNRAS.504...16G}. In numerical simulations the uniform UVB intensity can be adjusted to such that the mean transmitted flux $\bar F$ of simulated \lya\ forest spectra matches that of real observation data. However, there are degeneracies since the gas density distribution and IGM temperature also affect $\bar F$. The thermal evolution of the IGM, while not directly sensitive to the {\em amplitude} of the UVB, is sensitive to the spectral shape since this determines the mean photoionization heating per recombination \citep{1997MNRAS.292...27H}. A more direct way to estimate the UVB intensity is based on detections of individual sources, i.e SFGs (e.g. \citealt{2018ApJ...869..123S, 2012ApJ...751...70V}) and quasars (e.g. \citealt{2009ApJ...692.1476C,2019A&A...632A..45R}). Ionizing photons produced by massive stars or quasars could propagate beyond their local {H\,{\sc ii}} region, enlarging ionized bubbles and escaping unimpeded. By measuring the escaping fraction and mean free path (MFP) of the ionizing photons in addition to the source luminosity functions, the ionizing background intensity could be estimated. 

The spatial fluctuations of IGM properties such as temperature \citep{2011MNRAS.415..977M,2018MNRAS.477.5501K} as well as UVB intensity \citep{2005MNRAS.360.1471M,2010ApJ...713..383W,2009MNRAS.400.1461M,2018MNRAS.473..560D}, could perturb the \lya\ forest observable in corresponding scales. As implicated by the early work of \cite{1999ApJ...520....1C}, the UVB fluctuation could have more important contribution to the fluctuation of \lya\ forest power spectrum in the limit of very large scale. Previous works built analytical models to quantify the UVB fluctuation \citep{2014PhRvD..89h3010P,2017MNRAS.472.2643S, 2014MNRAS.442..187G,2019MNRAS.482.4777M}. Recent works show that it may affect the inferred bias of Ly$\alpha$-emitting galaxies \citep{2022MNRAS.516..572M}. Also, the effect on large scale structure could complicate the constraints on cosmological parameter such as neutrino mass $m_{\nu}$, scaler spectral index $n_{\rm s}$ and non-Gaussianity $f_{\rm NL}$ in post-reionization large scale structure surveys \citep{2019MNRAS.485.5059U}. Dedicated to measure the comoving 100 $\rm h^{-1}\,Mpc$ baryonic acoustic osillation (BAO) feature in the correlation function, Baryonic Oscillation Spectroscopic Survey \citep[BOSS;][]{2013AJ....145...10D} has enabled the measurement of \lya\ forest correlation function on large scales comparable to the mean free path of ionizing photons. As a successor of BOSS, the Dark Energy Spectroscopic Instrument \citep[DESI;][]{2016arXiv161100036D} will provide many more quasar sightlines and extend the redshift range with dense sampling, enhancing the viability of investigating astrophysical phenomena in large scales. Furthermore, sourced by discrete SFGs and quasars, the scale-dependent feature of UVB could shed light on the scale-dependent distribution of these source populations and how it traces the underlying matter distribution at large scale \citep{2005MNRAS.356..596M}. 

In this work, we consider the scale-dependent ionizing background fluctuation in light of discrete source distribution at large scales, based on the modeling of this effect in \cite{2014PhRvD..89h3010P}. We take the UV source distribution into account in the biasing model and show its effect on the large scale feature of \lya\ flux power spectrum and the cross-spectrum of the \lya\ forest with a galaxy survey. The cross-correlation of multiple tracers can reduce cosmic variance since both the \lya\ forest and the galaxies trace the same underlying structure \citep{2009PhRvL.102b1302S,2009JCAP...10..007M,2021PhRvD.104h3501O}. Moreover, it breaks the degeneracy at large scales between the effective bias $b_j$ and mean number density $\bar n_j$ of ionizing sources (although not the degeneracy with the mean free path of ionizing photons; see \citealt{2014ApJ...792L..34P}). We use the Fisher matrix framework to explore this in the context of a specific pair of surveys: the \lya\ forest observations ongoing with DESI, and the [\ion{O}{III}] emission line survey planned with the Nancy Roman Space Telescope \citep{2015arXiv150303757S, 2022ApJ...928....1W}. We show that DESI-Roman cross-correlation could shed light on the origin of UVB photons and this power could be improved by increasing the overlapping survey volume of the two surveys.

This paper is structured as follows. In Section~\ref{sec:formalism} we present our power spectrum expressions and Fisher matrix framework. In Section~\ref{sec:survey}, we layout the realistic survey parameters and the choice of fiducial model parameter values  priors implemented. The results are shown in Section~\ref{sec:results} and conclusions are summarized in Section~\ref{sec:conclusion}.

Throughout this work, we use cosmological parameters from the Planck 2015 ``TT+TE+EE+lowP+lensing+ext'' \citep{2016A&A...594A..13P}: $\Omega_m=0.3089$, $\Omega_{\Lambda}=0.6911$, $\Omega_bh^2=0.02230$, $H_0=67.74\,{\rm km\,s^{-1}\,Mpc^{-1}}$, and $n_s=0.9667$.

\section{Cross-correlation of Lyman-$\alpha$ forest and galaxies}
\label{sec:formalism}

\subsection{Biasing Models}

A cosmological biasing model quantifies how certain observables trace the fluctuation of matter density in the Universe as well as other physical effects by a factor of corresponding bias coefficients. For the \lya\ forest the fluctuation of transmission flux is defined as
\begin{equation}
    \delta_{\rm F}=\frac{F}{\overline{F}(z)}-1,
\end{equation}
where $F$ is the fraction of quasar flux transmission and $\overline{F}(z)$ is its mean as a function of redshift $z$.

 \begin{table*}
    \centering
    \caption{\label{tab:symbol}Symbols used in this work.}
    \begin{tabular}{c|c}
    \hline
    \hline
    Symbol &  Definition \\
    \hline
    \multicolumn{2}{c}{\textit{Quantities included in Fisher matrix analysis as parameter}}\\ \
    $b_{\rm HI,u}$  & Bias of \HI\ in uniform UVB limit \\
    $b_{\rm clump}$ & Bias of \HI\ clumps \\
    $b_{\rm j}$     & Effective bias of ionizing sources\\
    $b_{\rm F\delta}$ & Bias of \lya\ flux to matter overdensity \\
    $b_{\eta}$      &  Bias of \lya\ flux to peculiar velocity gradient \\
    $b_{\rm g}$     & Bias of galaxy \\
    $\beta_{\rm r}$ & Dimensionless recombination radiation\\
    $\overline{n}_{\rm j}$ & Effective mean ionizing source number density\\
    $\kappa_{\rm HI}$ & Physical Lyman-limit opacity from IGM \HI\ and self-shielded clumps\\
    $p_{\rm clump}$ &  Fraction of self-shielded clumps opacity within $\kappa_{\rm HI}$, $\kappa_{\rm clump}/\kappa_{\rm HI}$\\
    \hline
    \multicolumn{2}{c}{\textit{ Other model quantities}}\\
    $b_{\rm HI}$    & Scale-dependent \HI\ bias \\
    $b_{\rm j,eff}$ & Effective bias of sources including recombination\\
    $\kappa_{\rm tot}$ & Effective total Lyman-limit opacity\\
    $\beta_{\rm clump}$ & Fraction from self-shielded clumps opacity of total opacity, $\kappa_{\rm clump}/\kappa_{\rm tot}$ \\
    $\beta_{\rm HI}$ & Fraction from IGM opacity of total opacity, $\kappa_{\rm HI}/\kappa_{\rm tot}$ \\
    $\beta_{\rm z}$ & Fraction from  effective redshifting opacity of total opacity\\
    $\beta_{\rm V}$ & Fraction from  effective volume dilution opacity of total opacity\\
    
    $\Tilde{\delta}_{\rm SN,j}$ & Shot noise of source number density \\
    $\Tilde{\delta}_{\rm SN,g}$ & Shot noise of galaxy number density \\
    $\overline{n}_{\rm Ly\alpha}$ &  Mean area number density of quasars in \lya\ forest survey\\
    
    $\overline{n}_{\rm g}$ & Effective mean galaxy number density\\ 
    \hline
    \end{tabular}
\end{table*}

The vanilla Lyman-$\alpha$ forest biasing model consider flux fluctuation at linear order with redshift space distortion effect:
 \begin{equation}
 \tilde\delta_{\rm F}({\bmath k})=b_F(1+\beta_{\rm F}\mu^2)\tilde\delta_{\rm m}({\bmath k}),
 \end{equation}
 where $b_{\rm F}$ is the usual Lyman-$\alpha$ flux bias, $\delta_{\rm m}$ is the matter overdensity, $\beta_{\rm F}$ is the redshift distortion parameter and $\mu=\cos{\theta}=k_{\parallel}/k$ the angle between the Fourier wavevector and the line of sight. Since Lyman-$\alpha$ transmission is not conserved in the real-to-redshift space conversion, the usual relation \citep{1987MNRAS.227....1K} $\beta = f/b$ (where $f$ is the growth rate of structure and $b$ is the bias) does not apply and $\beta_F$ (or equivalently the biasing coefficient $b_\eta = b_F\beta_{\rm F}/f\neq 1$) must be treated as a parameter. In this work, we would like to study the UVB source fluctuation by its imprints on \lya\ flux power spectrum, particularly on its large scale feature. We adapt the \HI\ biasing model built in \cite{2014PhRvD..89h3010P} and summarize symbols throughout this work in Table~\ref{tab:symbol}. (Although the parametrization is somewhat different, the \citealt{2014MNRAS.442..187G} model includes a source bias and shot noise and has many of the same functional dependencies, and we expect that similar results could be obtained using that framework.)  The \HI\ overdensity in Fourier space in this model is written as
\begin{equation}
    \tilde{\delta}_{\rm HI}({\bmath k}) = b_{\rm HI}(k)\tilde{\delta}_{\rm m}({\bmath k})-\frac{[1-\beta_{\rm HI}\beta_r]S(k)}{1-\beta_{\rm HI}S(k)}\tilde{\delta}_{\rm SN}({\bmath k}).
\end{equation}
The characteristic function $S(k)$ is the normalized Fourier transform of the flux profile from an ionizing source including both the inverse square law and exponential attenuation:
\begin{eqnarray}\label{eq:Sk}
    S(k) &=&
\int_{{\mathbb R}^3} a\kappa_{\rm tot} {\rm e}^{-a\kappa_{\rm tot}r}\frac1{4\pi r^2} {\rm e}^{{\rm i}{\bmath k}\cdot{\bmath r}}\,{\rm d}^3{\bmath r}
\nonumber \\ &=&
\frac{a\kappa_{\rm tot}}{k}\arctan{\frac{k}{a\kappa_{\rm tot}}},
\end{eqnarray}
where the scale factor $a$ converts from physical to comoving attenuation coefficient. The scale-dependent \HI\ bias $b_{\rm HI}$ is 
\begin{equation}
\begin{aligned}
    b_{\rm HI}(k)
    &= \frac{b_{\rm HI,u}-b_{\rm j,eff}S(k)}{1-\beta_{\rm HI}S(k)}\\
    &= \frac{b_{\rm HI,u}}{1-\beta_{\rm HI}S(k)}\\
    & \quad-\frac{[(1-\beta_{\rm HI}\beta_{\rm r})b_j-\beta_{\rm clump}b_{\rm clump}+\beta_{\rm HI}\beta_{\rm r}b_{\rm HI,u}]S(k)}{1-\beta_{\rm HI}S(k)},
\end{aligned}
\end{equation}
where we substitute the effective source bias $b_{\rm j,eff}$ as given in \citet{2014PhRvD..89h3010P}:
\begin{equation}
b_{\rm j,eff}=(1-\beta_{\rm HI}\beta_{\rm r})b_j-\beta_{\rm clump}b_{\rm clump}+\beta_{\rm HI}\beta_{\rm r}b_{\rm HI,u}.
\end{equation}

We incorporate the \lya\ flux bias and redshift-space distortion into this \HI\ model to get a \lya\ flux biasing model. The \lya\ flux fluctuation in Fourier space is written as
\begin{eqnarray}\label{eq:delta_flux}
\tilde{\delta}_{\rm F}({\bmath k})
&=& \frac{b_{\rm F\delta}}{b_{\rm HI,u}}\tilde{\delta}_{\rm HI}({\bmath k}) + b_{\eta}\mu^2f\tilde{\delta}_{\rm m}({\bmath k}) \nonumber \\
&=& \left[\frac{b_{\rm HI}(k)b_{\rm F\delta}}{b_{\rm HI,u}}+b_{\eta}\mu^2f\right]\tilde{\delta}_{\rm m}({\bmath k})
\nonumber \\ && ~~~~
-\frac{b_{\rm F\delta}}{b_{\rm HI,u}}\frac{[1-\beta_{\rm HI}\beta_r]S(k)}{1-\beta_{\rm HI}S(k)}\Tilde{\delta}_{\rm SN,j}.
\end{eqnarray}
On small scales($k\gg\kappa_{\rm tot}$), S(k) asymptotes to zero such that the model reduces to the uniform UVB background assumption since $b_{\rm HI}(k) \rightarrow b_{\rm HI,u}$. Conversely on large scales, $S(k\ll\kappa_{\rm tot})\rightarrow 1$, and so the bias becomes
\begin{equation}
\lim_{k\rightarrow 0}
\frac{b_{\rm HI}(k)b_{\rm F\delta}}{b_{\rm HI,u}}
= \frac{1-b_{\rm j,eff}/b_{\rm HI,u}}{1-\beta_{\rm HI}} b_{\rm F\delta}.
\end{equation}
If $b_{\rm j,eff}>b_{\rm HI,u}$, this can even flip sign and become positive -- that is, ultra-large-scale overdensities can have increased rather than decreased transmitted flux $F$, because they have a more intense UVB. However with the scale cuts ($k_{\rm min}$) we use in this paper, we usually do not enter this regime. Also in the case of ionizing background fluctuations, the shot noise term becomes important on large scales rather than small scales because of the scale dependence of $S(k)$.
Another feature is that this bias diverges as $k\rightarrow 0$ if $\beta_{\rm HI} = 1$ (see commentary in the introduction of \citealt{2019MNRAS.482.4777M}). This is the limit where only diffuse intergalactic gas contributes to removal of ionizing photons, with the ``clumpy'' absorbers and redshifting turned off. The divergence occurs because in such a model, there are either enough recombinations in the IGM to produce the \ion HI required to absorb the ionizing photons contributed by sources (in which case the IGM recombines) or not (in which case the UVB simply builds up with time, approaching $\infty$ in the time-steady limit due to Olbers' paradox). A model with $\beta_{\rm HI}=1$ and finite UVB, balanced exactly on this phase transition, has an infinite coefficient for response to perturbations in the time-steady limit. In our fiducial model $\beta_{\rm HI} = 0.59$; in general, in the presence of the clumping and redshifting ``sinks'' for UVB photons, the divergent bias problem does not apply.

For the galaxies, since we are interested mainly in the large scales, we take the simple linear biasing model
\begin{equation}
    \tilde{\delta}_{\rm g}({\bmath k}) = (b_{\rm g}+f\mu^2)\tilde{\delta}_{\rm m}({\bmath k})+\tilde{\delta}_{\rm SN,g}({\bmath k}),
\end{equation}
where $f$ is the growth rate and $\delta_{\rm SN,g}$ is a galaxy shot noise term. Note that the ``tracer'' galaxy sample is not necessarily the same as the population of galaxies that is emitting the ionizing photons, so $b_{j,\rm eff}$ and $b_g$ are different parameters.

\subsection{Power spectra}

The auto-correlation power spectrum $P_{\rm F}(k,\mu)$ of the \lya\ flux transmission has three pieces: a biasing term coming from $\tilde\delta_m({\bmath k})$ that traces the matter power spectrum; an ionizing source shot noise term; and observational terms (themselves coming from the finite signal-to-noise ratio of quasar spectra, and the aliasing effect from sampling a finite density of sightlines).
The first two can be derived from Eq.~(\ref{eq:delta_flux}).
\begin{equation}
   \begin{aligned}
    P_{\rm F}(k,\mu) = & 
    \overbrace{b_{\rm F}^2(k,\mu)P_{\rm m}(k)}^{\text{\normalsize biasing, $P_{\rm F,b_{\rm HI}}$}}
    +\overbrace{\frac{1}{\overline{n}_{\rm j}}\left[\frac{b_{\rm F\delta}}{b_{\rm HI,u}}\frac{(1-\beta_{\rm HI}\beta_{\rm r})S(k)}{1-\beta_{\rm HI}S(k)}\right]^2}^{\text{\normalsize source shot noise, $P_{\rm F,SN}^{\rm j}$}} \\
    + & \underbrace{\frac{1}{\overline{n}_{\rm Ly\alpha}}\left[P_{\rm F}^{\rm noise}+P_{\rm F}^{\rm 1D}\right]}_{\text{\normalsize observation noise, $P_{\rm F,SN}^{Ly\alpha}+P_{\rm F,alias}$}},\label{eq:pf}
\end{aligned} 
\end{equation}
where $P_{\rm m}$ is the matter power spectrum calculated by CLASS \citep{2011JCAP...07..034B}. The effective flux bias $b_{\rm F}$ is given by
\begin{equation}
b_{F}(k,\mu) \equiv \frac{b_{\rm HI}(k)b_{\rm F\delta}}{b_{\rm HI,u}}+b_{\eta}\mu^2f.
\end{equation}
The 1-dimensional noise power spectrum $P_{\rm F}^{\rm noise}$ \citep{2011MNRAS.415.2257M} is
\begin{equation}\label{eq:p_noise}
    P_{\rm F}^{\rm noise} = [S/N]^{-2}\frac{\Delta\lambda_{\rm obs}}{\lambda_{\rm Ly\alpha}}\frac{c}{H(z)},
\end{equation}
where $S/N$ is the signal noise in $\Delta\lambda_{\rm obs} = 0.8$ \AA\ pixels,  the rest-frame \lya\ wavelength is $\lambda_{\rm Ly\alpha} = 1216$ \AA, $c$ is the speed of light, and $H(z) = H_0\sqrt{\Omega_{m}(1+z)^3+\Omega_{\Lambda}}$.
For the aliasing contribution $P_{\rm F}^{\rm 1D}$ we use the empirical function fitted with the BOSS dataset and likelihood method in \cite{2013A&A...559A..85P}. Note that \cite{2020MNRAS.497.4742K} modifies the fitting function with a Lorentzian decay to better fit $P_{\rm 1D}$ in the high-$k$ regime ($k \sim 37 \rm h\,Mpc^{-1}$). But for k range considered in this work the high-$k$ correction will not significantly impact the results, so for our purpose the empirical function in \cite{2013A&A...559A..85P} is valid. 

The auto-correlation power spectrum for the galaxy survey $P_g(k,\mu)$ is
\begin{equation}\label{eq:pg}
    P_{\rm g}(k,\mu) = (b_{\rm g}+f\mu^2)^2P_{\rm m}(k)+\frac{1}{\overline{n}_{\rm g}},
\end{equation}
where $P_{\rm m}$ is the matter power spectrum, and $\overline{n}_{\rm g}$ is the effective source density in the galaxy survey. The cross-correlation of the Lyman-$\alpha$ forest and galaxy survey $P_{\times}(k,\mu)$ can be written as
\begin{equation}\label{eq:pc}
    P_{\times}(k,\mu) = (b_{\rm F}(k,\mu)+b_{\eta} f\mu^2)(b_{\rm g}+f\mu^2)P_{\rm m}(k).
\end{equation}
Note that by crossing the two different tracers, the shot noise terms in auto-correlations have been removed. (The shot noise term in the UVB source $1/\overline n_{\rm j}$ would be dominated by quasars since they are rare and bright. If we had used quasars as the cross-correlation tracer, then there would be a common shot noise term that depends on the overlap of the UVB quasars and the tracer quasars, described by an additional parameter; see \citealt{2014PhRvD..89h3010P}.) For two UVB source parameters $b_{\rm j}$ and $\overline{n}_{\rm j}$, the cross-correlation power spectrum will only be sensitive to the former. So the cross-correlation measurement could help break degeneracy between $b_{\rm j}$ and $\overline{n}_{\rm j}$.

\subsection{Fisher matrix}

We derive the detection sensitivity of cross-correlation between \lya\ and galaxy surveys to the scale-dependent UVB model under the frame of Fisher matrix formalism. Given data vector $\Vec{D} = (\delta_{\rm g},\delta_{\rm F})$ and parameter vector $\Vec{p}$, the covariance matrix is 
\begin{equation}
{\mathbfss C}(k,\mu)=
    \begin{pmatrix}
    P_{\rm g}(k,\mu) & P_{\times}(k,\mu) \\
    P_{\times}(k,\mu) & P_{\rm F}(k,\mu)
    \end{pmatrix}.
\end{equation}
The Fisher matrix for the measurement of a single mode is 
\begin{equation}
    F_{ij}=\frac{1}{2}{\rm Tr}\left[{\mathbfss C}^{-1}\frac{\partial \mathbfss C}{\partial p_i}{\mathbfss C}^{-1}\frac{\partial \mathbfss C}{\partial p_j}\right].
\end{equation}
Summing over all available modes in the comoving survey volume $V_{\rm sur}$, the Fisher matrix is
\begin{equation}
F_{\rm V}\simeq\frac{V_{\rm sur}}{4\pi^2}\int_{k_{\rm min}}^{k_{\rm max}}k^2\,{\rm d}k\int_{-1}^1 {\rm d}\mu\, F(k,\mu),
\end{equation}
where $V_{\rm sur} = \frac13\Omega[D_{\rm A}^3(z_{\rm max}) - D_{\rm A}^3(z_{\rm min})]$ is the comoving survey volume in the flat Universe ($\Omega_{\rm K}$ = 0) we assume in this work, $D_{\rm A}$ is the comoving angular diameter distance.

The Fisher matrix calculation is sensitive both to the small-scale and large-scale cutoffs in the $k$ integral, since linear bias parameters are best constrained from the large number of modes at high $k$ but the ionizing background fluctuation parameters have the largest effect at low $k$. We choose $k_{\rm max} = 0.2h\,{\rm Mpc}^{-1}$ as the fiducial linearity cut-off. We set a fiducial large-scale cutoff $k_{\parallel,\rm min}$ by requiring one wavelength across the redshift shell we are using,
\begin{equation}
    k_{\parallel,\rm min} = \frac{2\pi}{D_{\rm A}(z_{\rm max})-D_{\rm A}(z_{\rm min})}.
\end{equation}
In principle, the survey also has a finite extent in the transverse direction, which could in principle lead to a survey area-dependent $k_{\perp,\rm min}$. However, the angular size of a survey that corresponds to a ``cube'' in redshift space is $[c\Delta z/D(z)H(z)]^2$\,sr, which corresponds to 37 deg$^2$ for our fiducial redshift range. Since all of the planned surveys have contiguous area that is much larger than this, we have included only the radial finite size effects.

In the fiducial calculation of Fisher matrix, we have a redshift bin centered at $z=2.4$ and spanning $2.15<z<2.65$, 20 logarithmically spaced $k$-bins from 0.01497 to 0.2$h$ Mpc$^{-1}$, and $\mu$-bins with $|\mu|\geq \mu_{\rm min}=k_{\parallel, \rm min}/k$ for each k and $\Delta\mu=0.1$.

The ongoing DESI 3D correlation function analysis uses a cut in real space, $r_{\rm max} = 180 h^{-1}\,$Mpc (as was done in the eBOSS analysis; \citealt{2020ApJ...901..153D}); we note that $\pi/r_{\rm max} \approx 0.017h\,$Mpc$^{-1}$, which is not far from our fiducial cut. If one were to do a fit in correlation function space, it is still necessary to take into account that the continuum fitting procedure effectively high pass filters the Lyman-$\alpha$ forest, thereby suppressing any smooth component in the correlation function; previous analyses have taken this into account by applying this filtering to the theoretical correlation function before comparing with the data. This filtering is carried out by the $D_{AB}$ matrix in the BOSS/eBOSS analyses \citep{2017A&A...603A..12B, 2019A&A...629A..85D, 2020ApJ...901..153D}. While $D_{AB}$ describes the full filtering, we can get a simple estimate by noting that if we have a Fourier mode with some radial wave number $k_\parallel$, and project out the best-fit polynomial of order $N$ over a skewer of length $L$, the variance is reduced by a factor 
\begin{equation}
f_D = 1-\sum_{\ell=0}^N (2\ell+1) \left[j_\ell\left(\frac{k_\parallel L}2\right)\right]^2,
\label{eq:fD}
\end{equation}
where $j_\ell$ represents the spherical Bessel function. (A proof of this result is provided in Appendix~\ref{app:fd}.) The recent eBOSS analysis has $N=1$ and skewers of mean length $L=222h^{-1}\,$Mpc (mean wavelength range $\Delta\log_{10}\lambda = 0.049$ per skewer based on the pixel counts Table 1 of \citealt{2020ApJ...901..153D}), leading to $f_D = 0.8$ at $k_\parallel=0.031h\,$Mpc$^{-1}$; $f_D=0.1$ at $k_\parallel=0.014h\,$Mpc$^{-1}$; and $f_D=0.01$ at $k_\parallel=0.0076h\,$Mpc$^{-1}$. It is evident that the filtering removes the very large low-$k$ peak contributed from the shot noise term (top panel of Fig.~\ref{fig:ps}). However a DESI$\times$Roman analysis will presumably use a future version of the continuum fitting and there may be some improvement in the handling of the largest-scale modes \citep{2017A&A...603A..12B}. We are not certain how much improvement to expect, so in what follows we will consider forecasts where the large-scale cut is varied.

\subsection{Combining surveys}

In our calculation, the final Fisher matrix summing over information from cross-correlated survey strategy is built from 5 basic Fisher matrices calculated by the similar machinery described above: 
\begin{itemize}
    \item $F_{\rm Ly\alpha\times Gal, fid}$: The information matrix from \lya\ and galaxy survey cross-correlated area with fiducial galaxy survey strategy. In our case, fiducial and deep galaxy survey strategies differ by deep survey has longer exposure time and thus larger number of $\overline{n}_{\rm g}$ and smaller Poisson noise for power spectrum. We describe the galaxy survey strategy in detail in \S\ \ref{subsec:strategy}.
    
    \item $F_{\rm Ly\alpha\times Gal, deep}$: The information matrix from \lya\ and galaxy survey cross-correlated area with deep galaxy survey strategy.
    
    \item $F_{\rm Ly\alpha}$: The information matrix from \lya\ forest survey only area. We implement it by setting $\overline{n}_{\rm g}=10^{-12} \,\rm h^3\,Mpc^{-3}$ in Eq.~(\ref{eq:pg}) to bury information from galaxy survey under noise such that effectively only \lya\ forest provides information about cosmology.
    
    \item $F_{\rm Gal,fid}$: The information matrix from galaxy survey only area with the fiducial galaxy survey strategy. We implement this by setting $P_{\rm F}^{\rm noise}=10^{12}h^{-1}\, \rm Mpc$ in Eq. \ref{eq:pf} to cover the information from \lya\ forest survey. 
    
    \item $F_{\rm Gal,deep}$: The information matrix from galaxy survey only area with the deep galaxy survey strategy.
    
\end{itemize}
The total information matrix from the combined surveys is
\begin{eqnarray}
 F_{\rm tot, fid(deep)} \!\!\!\! &=& \!\!\!\! F_{\rm Ly\alpha\times Gal, fid(deep)}(\Omega_{\rm cross}) + F_{\rm Ly\alpha}(\Omega_{\rm Ly\alpha}) \nonumber \\
 && + F_{\rm Gal,fid
    (deep)}(\Omega_{\rm Gal}) + F_{\rm prior}
\end{eqnarray}
where $\Omega_{\rm cross}$ is overlapping survey area solid angle, $\Omega_{\rm Ly\alpha}$ is the \lya\ forest survey only area angle, and $\Omega_{\rm Gal}$ is the galaxy survey only area angle. We include a prior information matrix $F_{\rm prior}$, with choices described in Section \ref{subsec:parameter}.

\section{Survey and model Parameters}
\label{sec:survey}

\subsection{DESI and Roman}\label{subsec:strategy}

In this work, we benchmark our model by producing a realistic \lya\ flux and galaxy cross-correlation between the \lya\ Forest measurement of DESI and the high latitude spectroscopic survey (HLSS; \citealt{2022ApJ...928....1W}) of Roman at the overlapping redshifts $2.15<z<2.65$.

For Roman, we consider both a ``fiducial’’ and ``deep’’  [\ion{O}{III}] emitter number density. Both of these densities are based on the Exposure Time Calculator v19 \citep{2012arXiv1204.5151H}. We assumed the average of the luminosity functions from grism \citep{2013ApJ...779...34C, 2015ApJ...811..141M} and narrow-band \citep{2015MNRAS.452.3948K} surveys (model 1992 in the ETC). We used an updated throughput table based on the Phase C payload design\footnote{Accessible at the Roman Space Telescope website:  \href{https://roman.gsfc.nasa.gov/science/RRI/Roman\_effarea\_20210614.xlsx}{https://roman.gsfc.nasa.gov/science/RRI/Roman\_effarea\_20210614.xlsx}} and a signal-to-noise ratio threshold of 6. The ``fiducial’’ number density corresponds to the reference exposure count and duration of $6\times 297$ s; and the ``deep’’ number density corresponds to observations twice as long ($6\times 594$ s).

For the DESI \lya\ forest survey, we quote quasar number density in Table 2.7 of \cite{2016arXiv161100036D} for $\overline{n}_{Ly\alpha}$ in Eq.~(\ref{eq:pf}). We use the typical signal-to-noise ratio for DESI quasar spectra of $S/N=2$ per re-sampled spectral pixel of width $\Delta \lambda_{obs}=0.8$ \AA. DESI is conducting its survey over 14,000 square degrees, split among the North and South Galactic caps; the declination range in the North Galactic Cap is $\delta>-8.2^\circ$, and that in the South Galactic Cap is $-18.4^\circ <\delta < +30^\circ$. By drawing $4\times 10^6$ Monte Carlo samples from the sky, we find that 40,274 fall into the overlap of the reference Roman footprint and the DESI survey area, thus under the Reference survey plan they would overlap for $\Omega \approx 415.4 \, {\rm deg}^2$ in the sky. We show in Sec.~\ref{sec:results} how varying survey strategy would improve the sensitivity to large-scale ionizing background parameters. 

\subsection{Parameter fiducial value and priors}
\label{subsec:parameter}

\begin{table*}
\caption{Summary of fiducial value, priors and forecast $1\sigma$ sensitivity with the fiducial survey strategy at $z=2.4$ and overlap area of $\Omega_{\rm cross}=415$ deg$^2$. We show the forecast for our fiducial $k$-range, and for two extended $k$-ranges: one where $k_{\rm max}$ is doubled to $2\times 0.2h$\,Mpc$^{-1}$, and one where $k_{\parallel,\rm min}$ is halved to $\frac{1}{2}\times 0.01497h\,{\rm Mpc}^{-1}$. In the last two columns, we show the improvement in constraint that could be obtained with this extension of the $k$-range.}
\begin{tabular}{cccccccrr}
\hline
\hline
 Parameter & Unit & Fiducial Value & Prior $1\sigma$ & \multicolumn3c{Forecast $1\sigma$} & \multicolumn2c{Improvement}
 \\
 ~ & ~ & ~ & ~ & Fiducial & $k_{\rm max}${\tt *=}2 & $k_{\parallel,\rm min}${\tt/=}2 & $k_{\rm max}${\tt *=}2 & $k_{\parallel,\rm min}${\tt/=}2 \\
\cmidrule(lr){5-7}\cmidrule(lr){8-9} 
 \hline
  \multicolumn9c{\textit{Constrained Parameters}}\\
$b_{\rm j}$ & - & 3 & - & 0.456 & 0.423 & 0.424
& 7.2\% & 7.0\% \\
$1/\overline{n}_{\rm j}$ & $h^{-3}$\,Mpc$^3$ & 1.36$\times10^4$ & - & 1.665$\times10^4$ & 1.437$\times10^4$ & 0.895$\times10^4$
& 13.4\% & 44.1\% \\
$b_{\rm F\delta}$ & - & -0.1116 & - & 0.00099 & 0.00074 & 0.00078
& 25.2\% & 21.3\% \\
$b_{\eta}$ & - & -0.1594 & - & 0.00143 & 0.00098 & 0.00122
& 31.5\% & 14.4\% \\
$b_{\rm g}$ & - & 2.4 & - & 0.00592 & 0.00387 & 0.00575
& 34.7\% & 2.9\% \\ \multicolumn9c{\textit{Unconstrained Parameters}} \\
 $p_{\rm clump}$ & - & 0.09 & 0.02 & 0.02 & 0.02 & 0.02\\
 $\kappa_{\rm HI}$ & Mpc$^{-1}$ & 0.0068  & 0.00078 & 0.00078 & 0.00078 & 0.00078 \\
 $b_{\rm HI,u}$ & - & 1.6 & 0.2 & 0.2 & 0.2 & 0.2\\
 $\beta_{\rm r}$ & - & 0.3997 & 0.00796 & 0.00796 & 0.00796 & 0.00796 \\
 $b_{\rm clump}$ & - & 1.6 &  0.2 & 0.2 & 0.2 & 0.2\\
 \hline
\hline
\end{tabular}
\label{tb:params}
\end{table*}

\begin{table*}
\caption{Summary of parameter constraints improvement in DESI+Roman going from disjoint ($\Omega_{\rm cross}=0$) to maximal overlap ($\Omega_{\rm cross} = 2227$ deg$^2$).}
\begin{tabular}{lcccrccr}
\hline
\hline
Parameter & Unit & \multicolumn3c{Fiducial galaxy survey} & \multicolumn3c{Deep galaxy survey} \\
 \cmidrule(lr){3-5}\cmidrule(lr){6-8}
& & \multicolumn2c{Forecast 1$\sigma$} & Improvement & \multicolumn2c{Forecast 1$\sigma$} & Improvement \\
~ & ~ & Disjoint & Maximal & ~ & Disjoint & Maximal & ~ \\
\cmidrule(lr){3-4}\cmidrule(lr){6-7}
\hline
$b_{\rm j}$ & - & 0.653 & 0.410 & 37.2\% & 0.653 & 0.405 & 37.9\%  \\
$1/\overline{n}_{\rm j}$ & $h^{-3}$\,Mpc$^{-3}$ & $3.814\times10^4$ & $0.888\times10^4$ & 76.7\% & $3.814\times10^4$ & $0.795\times10^4$ & 79.2\% \\
$b_{\rm F\delta}$ & - & 0.00151 & 0.00083 & 45.0\% & 0.00151 & 0.00079 & 47.7\% \\
$b_{\eta}$ & - & 0.00230 & 0.00117 & 48.9\% & 0.00230 & 0.00112 & 51.1\% \\
$b_{\rm g}$ & - & 0.00594 & 0.00585 & 1.5\% & 0.00430 & 0.00423 & 1.6\% \\
\hline
\hline
\end{tabular}\label{tb:params_improve}
\end{table*}

\begin{figure}
    \centering
    \includegraphics[width=3in]{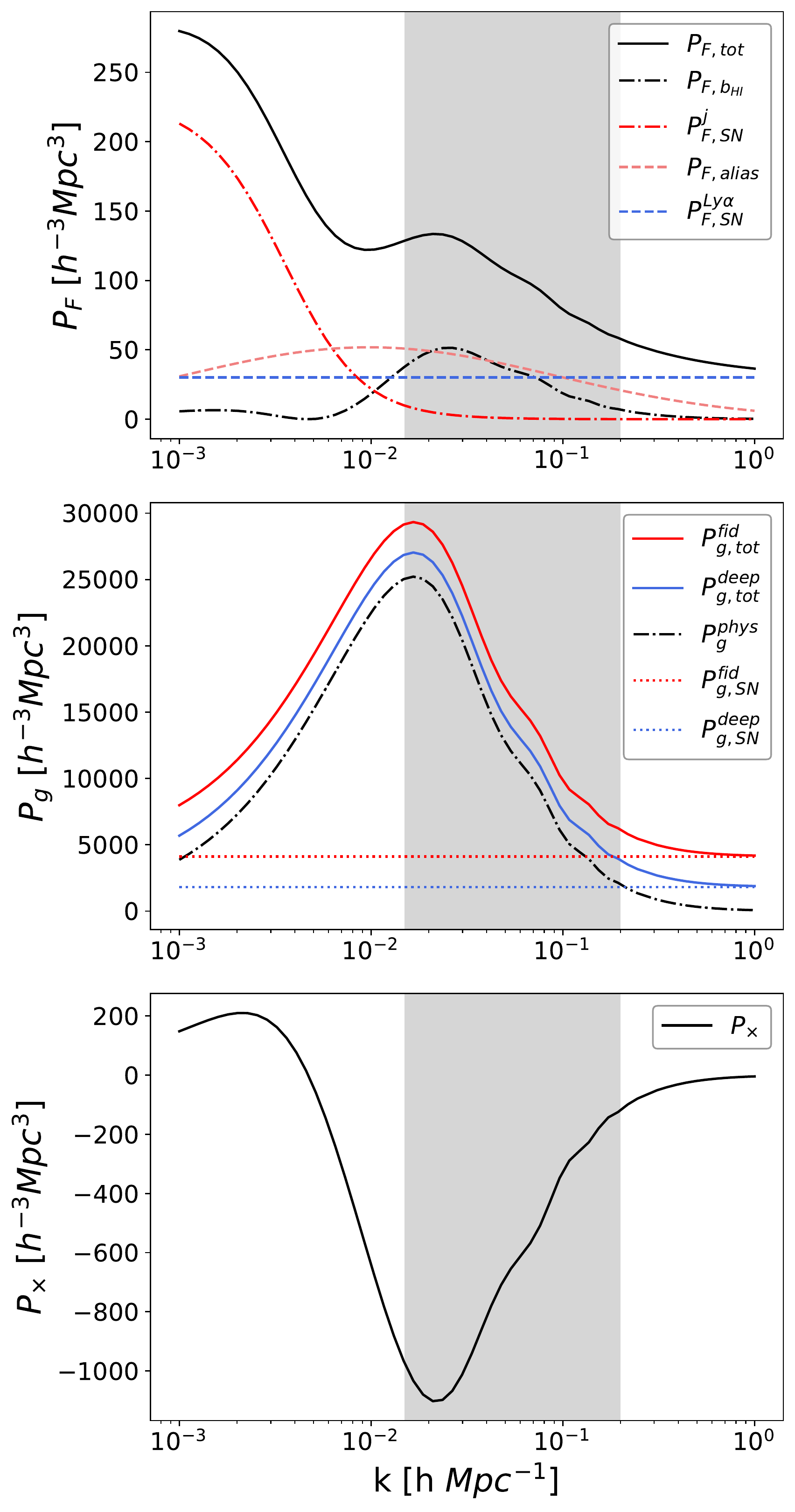}
    \caption{(\textit{Upper panel}) power spectrum of \lya\ forest flux. Black solid line is the total power spectrum; black dash dot line is the theoretical power spectrum take UVB source contribution into account; red dash dot line represents the contribution from UVB source number shot noise; the blue and coral dashed line represents the \lya\ shot noise from detector noise and alias term respectively. (\textit{Middle panel}) galaxy power spectrum with fiducial and deep survey strategy. Red solid and dotted line represents the total and shot noise power spectrum respectively for fiducial exposure time, while the blue lines corresponds to results with longer exposure time (deep survey). The black dash-dotted line is the theoretical galaxy power spectrum. (\textit{Lower panel}) cross-correlated power spectrum of \lya\ forest flux and galaxy survey. The grey blocks represents the available $k$ range in the search of surveys. 
    \label{fig:ps}}
\end{figure}

In this work, we explore the cross-correlation strategy constraints on the 10 model parameters as collected in the upper portion of Table~\ref{tab:symbol}.

Here we describe physical meaning, and the choice of fiducial values and prior uncertainties (if any) at $z=2.4$ for each parameter. Much of this builds on the choices in \cite{2014PhRvD..89h3010P}, but there are some updates.  We use the tabulated mean UVB estimated in \cite{2012ApJ...746..125H} for fiducial values as they enter into the calculation of $p_{\rm clump}$, $\beta_{\rm r}$, $b_{\rm HI,u}$.
\begin{itemize}
    \item $\kappa_{\rm HI} = (\beta_{\rm clump}+\beta_{\rm HI})\kappa_{\rm tot}=\overline{\sigma}_{\rm HI}n_{\rm HI}+\overline{\kappa}_{\rm clump}$: The physical Lyman-limit opacity, which is related to Lyman-limit photon mean free path $\lambda_{\rm mfp}$ by $\kappa_{\rm HI} = 1 /\lambda_{\rm mfp}$. We take a fiducial value $\lambda_{\rm mfp}=147\,\rm Mpc$ at $z=2.4$ as measured by \citet{2013ApJ...769..146R}.
    There is a strong degeneracy between $\kappa_{\rm HI}$ and the ionizing source parameters \citep{2014ApJ...792L..34P}, since $S(k)\propto k^{-1}$ over most of the observable range of $k$ and the scale and amplitude are exactly degenerate for a power law. Thus the incorporation of some external information on $\kappa_{\rm HI}$ is necessary.
    We take the uncertainties in the measurement of opacity in \cite{2013ApJ...769..146R}, which breaks the uncertainty $\sigma_{\kappa_{\rm HI}}$ into the opacity uncertainties for Lyman-limit system $\sigma_{\kappa,\rm LLS}$ and forest opacity $\sigma_{\kappa,\rm for}$:
    \begin{equation}
    \sigma_{\kappa_{\rm HI}}=\sqrt{\sigma_{\kappa,\rm LLS}^2+\sigma_{\kappa,\rm for}^2}\approx 0.00078\, \rm Mpc^{-1}\,.
    \end{equation}
    (The conversion to $h$\,Mpc$^{-1}$ occurs inside the Fisher code.)
    \item $p_{\rm clump} = \beta_{\rm clump}/(\beta_{\rm clump}+\beta_{\rm HI})$: The fractional opacity contributed by clumping regions within the total opacity of intergalactic \HI\ and clumps. To find a fiducial value of this parameter, as shown in \S\ {\sc ii}.A of \cite{2014PhRvD..89h3010P}, at first one needs to identify the critical \HI\ column density that distinguishes IGM and clumping regions by finding the cross point between curves of IGM \HI\ photoionization rate and the collisional ionization rate in clumps. Thereafter, an estimation of $p_{\rm clump}$ is made by checking the measured fraction of IGM contributed opacity for a certain \HI\ column density in Figure 10 of \cite{2013ApJ...769..146R}. The simple uniform-density 1D clump model in \cite{2014PhRvD..89h3010P} gives a column density $N\approx10/\overline{\sigma}_{\rm HI}\approx 2.6\times10^{18}\,{\rm cm}^{-2}$. We extend the analysis to a 2D clump model (see Appendix \ref{appendix:clump} for details of the modeling and estimation) in order to get an estimation of uncertainty for $p_{\rm clump}$. We find in the 2D model gives the column density a factor of 1/2, that is,  $N\approx1.3\times10^{18}\,{\rm cm}^{-2}$. Then the IGM should contribute opacity for around $89\% - 94\%$ so we choose 0.09 as the fiducial value for $p_{\rm clump}$ and 0.02 as the prior uncertainty.
    \item $b_{\rm HI,u}$: The bias of \HI\ in uniform UVB limit. We follow the discussion in \cite{2014PhRvD..89h3010P} for this value, while the \HI\ number density $n_{\rm HI}\propto \alpha(T)\rho^2$, the IGM equation-of-state $T\propto\rho^{\gamma-1}$ and recombination coefficient $\alpha(T)\propto T^{-0.7}$ (the exponent ranges from $-0.74$ at $T=8\times 10^3$ K to $-0.66$ at $T=2\times 10^4$ K; \citealt{1991A&A...251..680P}), this bias could be estimated by
    \begin{equation}
        b_{\rm HI,u} = \frac{\delta_{n_{\rm HI,u}}}{\delta_{\rho}}=2-0.7(\gamma-1),
    \end{equation}
    where $\gamma=1.6$ for a photo-heated IGM long after a reionization event (see \citealt{2016MNRAS.456...47M} for a thorough discussion), and thus giving a fiducial $b_{\rm HI,u}$ value of 1.6. We choose the prior uncertainty for this bias as $\sigma_{b_{\rm HI,u}}=0.2$ (i.e., $\sigma_\gamma=0.29$), given that \ion{He}{II} reionization could leave a significant imprint at this redshift \citep{2016MNRAS.460.1885U, 2018ApJ...865...42H, 2019ApJ...872...13W}.
    \item $\beta_{\rm r}$: The fraction of Lyman-limit photons from \HI\ recombinations. We use Eq. 18 of \cite{2014PhRvD..89h3010P} and forest temperature measurement in \cite{2011MNRAS.410.1096B} to estimate this value. The prior uncertainty is determined by $\sigma_{\beta_{\rm m}}=(\partial\beta_{\rm r}/\partial { T})\Delta T$.
    \item $b_{\rm clump}$: The bias of the clumps. Similar to \cite{2014PhRvD..89h3010P}, we assume $b_{\rm clump}=b_{\rm HI,u}$, $\sigma_{\rm clump}=\sigma_{\rm HI,u}$. The uncertainty in $b_{\rm clump}$ is degenerate with the uncertainty in $b_{\rm j}$ since they both enter through $b_{\rm j,eff}$.
    \item $b_{\rm j}$: The effective bias of photon source objects. This bias is an effective quantity taking emissivity contributed by multiple source populations into account
    \begin{equation}
        b_{\rm j}=\sum_{{\rm source}\,i}\frac{j_{{\rm 0,source}\,i}}{j_0}b_{{\rm j,source}\,i}
    \end{equation}
    where $j_0=\sum_{{\rm source}\,i}j_{0,{\rm source}\,i}$ is the total emissivity. The value of $b_{\rm j}$ depends on the underlying source populations. The main interest of this work is to explore the constraint on $b_{\rm j}$ by \lya\ and galaxy cross-correlation. We follow the default value $b_{\rm j}=3.0$ in \cite{2014PhRvD..89h3010P} assuming an average contribution from highly biased quasars and a range of galaxy luminosities. We leave it as a free parameter (formally, the prior $\sigma_{b_{\rm j}}=\infty$).
    \item $1/\overline{n}_{\rm j}$: The effective inverse of source number density, also shot noise contribution to power spectrum
    \begin{equation}
        \frac{1}{\overline{n}_{\rm j}}=\sum_{{\rm source}\, i} \left(\frac{j_{{\rm 0,source}\,i}}{j_0}\right)^2\frac{1}{\overline{n}_{{\rm source}\,i}}.
    \end{equation}
    We use the inverse number density since the flux power spectrum formula is smooth even as $1/\overline{n}_{\rm j}\rightarrow 0$; thus we expect the Fisher matrix approximation to to be much better using $1/\overline{n}_{\rm j}$ as a parameter than $\overline n_{\rm j}$.
    \cite{2014PhRvD..89h3010P} gives an approximate estimation of $ 2\times10^{-5}h^3\,{\rm Mpc}^{-3} < \overline{n}_{\rm j} < 10^{-4}h^3\,{\rm Mpc}^{-3}$. We begin with a model with $\overline{n}_{\rm j} = 5\times 10^{-5}h^3\,{\rm Mpc}^{-3}$, which translates to $1/\overline{n}_{\rm j} = 2\times 10^{4}h^{-3}\,{\rm Mpc}^{3}$. 
    However, a finite quasar lifetime would be expected to reduce the effective shot noise on large scales since over a finite light-travel time one would average over many realizations of the shot noise \citep{2014MNRAS.442..187G, 2019MNRAS.482.4777M}. We quantify this issue in Appendix~\ref{app:variability}; the factor $f_{\rm FL}$ by which the shot noise is suppressed depends on the quasar lifetime $t_{\rm Q}$, the scale $k$, and the angle to the line of sight $\mu$ (since with time dependent effects the light travel time to the observer is also important). At the largest scale we use ($k=0.015h$ Mpc$^{-1}$, $\mu=1$) and a quasar lifetime of $t_{\rm Q}=100$ Myr, the suppression factor is $f_{\rm FL}=0.55$. This could be smaller for even shorter quasar lifetimes, but is also larger if we consider larger $k$ (i.e., not exactly at our minimum value). Thus we take a suppression factor corresponding to $k=0.02h$ Mpc$^{-1}$, $\mu=1$, $t_{\rm Q}=100$ Myr, i.e., $f_{\rm FL}=0.68$, or an effective $1/\overline{n}_{\rm j} = 0.68\times 2 \times 10^4 h^{-3}\,{\rm Mpc}^3 = 1.36\times 10^{4}h^{-3}\,{\rm Mpc}^{3}$, as our fiducial value.
    We leave it as a free parameter.
    \item $b_{\rm F\delta}$: The bias of \lya\ flux to matter overdensity. The fiducial values are taken from the fiducial simulation of \citet{2015JCAP...12..017A}. We found that the statistical errors from the DESI survey are below the current theory uncertainty of the bias parameters \citep{2015JCAP...12..017A}, especially if we try to marginalize over the temperature-density relation parameters, so imposing a prior does not significantly improve the constraints. Thus we took the conservative option of leaving it as a free parameter.
    \item $b_{\eta}$: The bias of \lya\ flux to peculiar velocity gradient. Again the fiducial value is from the fiducial simulation of \citet{2015JCAP...12..017A} and we leave $b_\eta$ as a free parameter.
    \item $b_{\rm g}$: The galaxy bias. We set the fiducial value equal to 2.4 \citep{2014PhRvD..89h3010P} and regard it as a free parameter, expecting the auto-correlation in the galaxy survey to constrain it very well (at fixed cosmology).
\end{itemize}

\section{Results}
\label{sec:results}

\subsection{Power Spectra}

We display our results for the baseline DESI+Roman surveys for $P_{\rm g}$, $P_{\rm F}$, $P_{\times}$ in Eqs.~(\ref{eq:pg}--\ref{eq:pc}) at $z = 2.4$, $\mu=1/\sqrt{3}$ in Figure~\ref{fig:ps}.  We paint by grey shadow the available survey wavenumber region used in this work, i.e. $0.01497\leq k \leq 0.2\,h\,\rm Mpc^{-1}$. The \lya\ flux power spectrum breaks into contribution from 3 pieces: the contribution from the first two terms of Eq.~(\ref{eq:pf}) corresponding to physical fluctuations in the \lya\ transmission $P_{\rm F, b_{\rm HI}}$; the observational noise in the spectra $P_{\rm F}^{\rm noise}/\overline{n}_{\rm Lya}$; and the aliasing term $P_{\rm F}^{\rm 1D}/\overline{n}_{\rm Lya}$. Note that the cross-correlation power spectrum is negative at most scales because the magnitude of flux transmission is anti-correlated with the galaxy density. However, at extremely large scales, $P_{\times}$ becomes positive: overdensities of matter (hence overdensities in the tracer galaxies since $b_{\rm g}>0$) have higher ionizing photon emissivity and higher ionizing background. Therefore, in ultra-large-scale overdense regions, the \lya\ flux transmission is indeed reduced by the increased gas density, but at the same time is increased by the stronger UVB. The latter effect wins out at the largest scales where $S(k)>b_{\rm HI,u}/b_{\rm j,eff}$.

\begin{figure}
    \centering
    \includegraphics[width=3in]{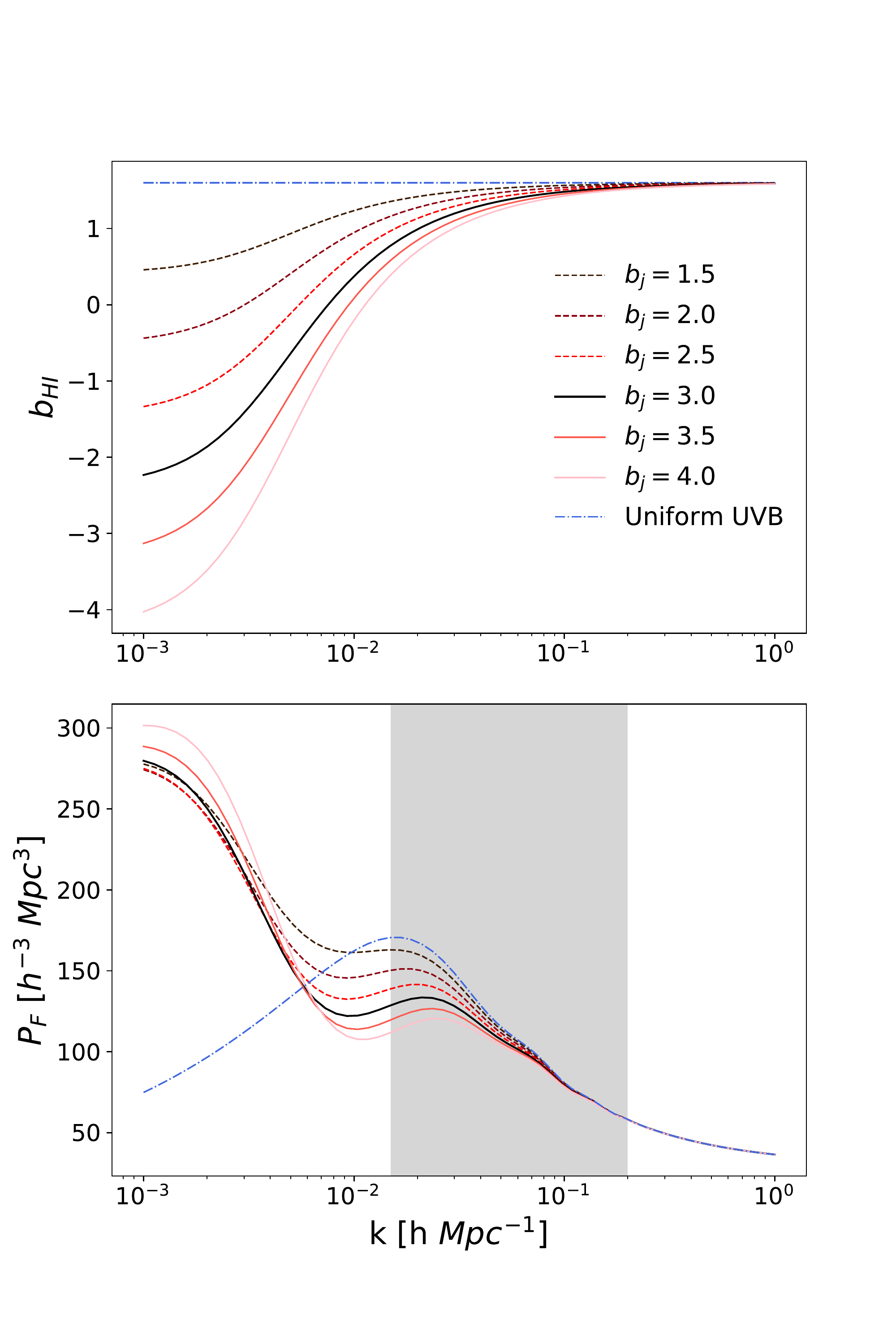}
    \caption{Variations of scale-dependent \HI\ bias $b_{\rm HI}(k)$ (upper panel) and \lya\ forest flux power spectrum $P_{\rm F}(k)$ with respect to varied effective ionizing source bias $b_{\rm j}$ at $z=2.4$, $\mu = 1/\sqrt{3}$. The blue dash-dot line shows the spatially uniform UVB scenario for comparison. At the upper panel stronger clustering of UVB sources (larger $b_{\rm j}$) suppresses the value of $b_{\rm HI}$ and shifts the zero point of it to smaller scales (larger k) because the enhancement of ionizing emissivity. At the lower panel the peak amplitude of $P_{\rm F}$ is suppressed and the dip is sharper also because UVB source radiation overcompensates the matter clustering.}
    \label{fig:pf_bj}
\end{figure}

We display the variation of scale-dependent \HI\ bias $b_{\rm HI}$ and $P_{\rm F}$ with respect to the effective source bias $b_{\rm j}$ in Figure~\ref{fig:pf_bj}. We also plot the uniform UVB scenario for comparison. At the scale of $k\approx\kappa_{\rm tot}=0.0100 \rm \,hMpc^{-1}$, the flux power spectra show a sharp dip since $b_{\rm HI}$ crosses zero. This feature also reflects the transition of transmitted flux from matter density dominated to UVB source dominated when goes to larger scales. At the far left side of Figure~\ref{fig:pf_bj}, as $k\rightarrow 0$, the matter clustering goes away entirely, $P(k)\rightarrow 0$, and the large scale power spectrum approaches a constant given by the source shot noise term in Eq.~(\ref{eq:pf}) with $S(k)\rightarrow 1$.

\begin{figure}
    \centering
    \includegraphics[width=3in]{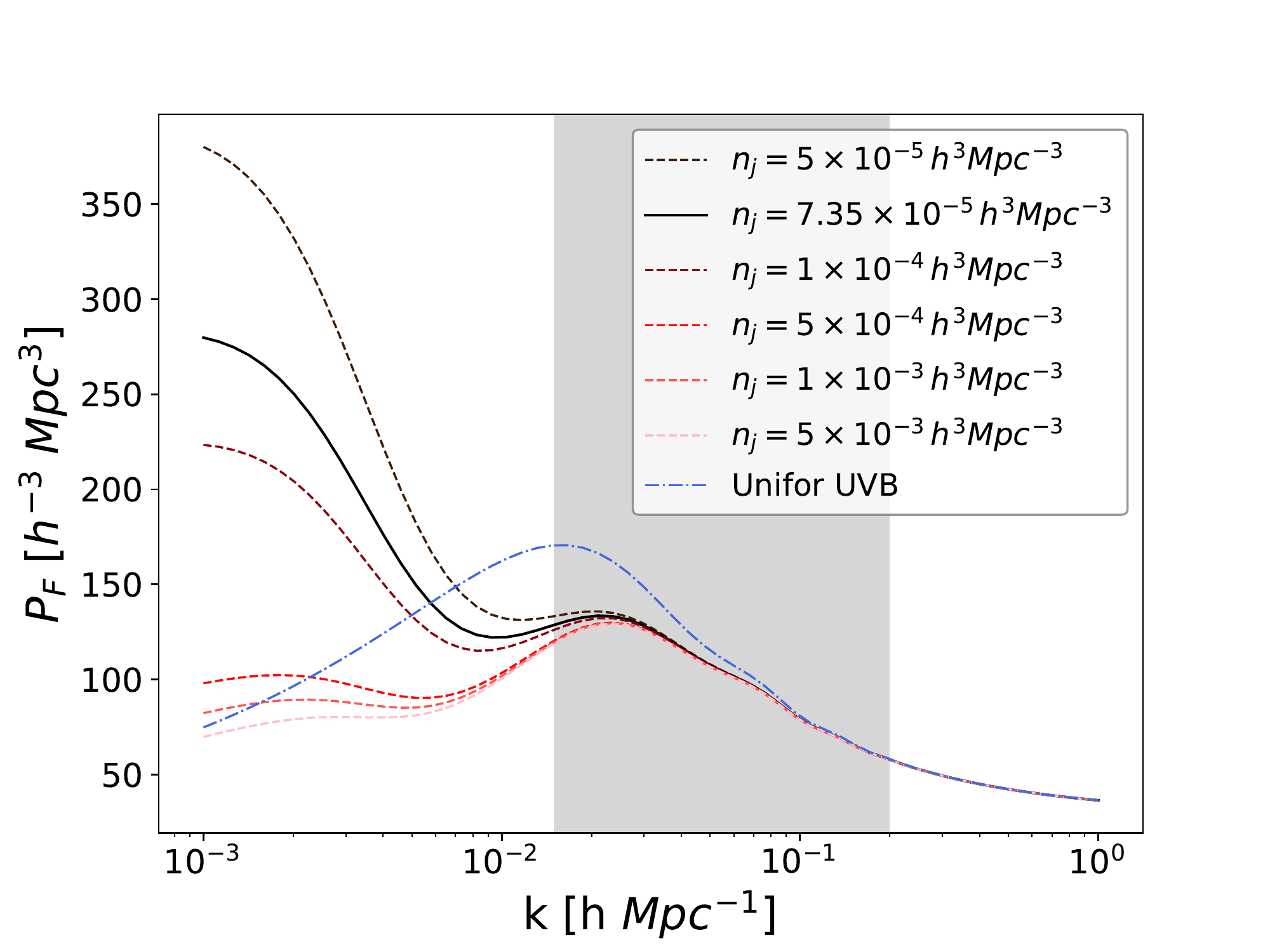}
    \caption{Variation of \lya\ forest flux power spectrum $P_{\rm F}(k)$ with respect to varied effective ionizing source mean number density $\overline{n}_{\rm j}$. The shot noise of ionizing source number gives rise to the large magnitude at large scales, as also shown in upper panel of Figure. \ref{fig:ps}.}
    \label{fig:pf_nj}
\end{figure}

Reflecting the Poisson fluctuation of the ionizing source number density, $\overline{n}_{\rm j}$ could impact the large scale $P_{\rm F}(k)$ as well. We show the variation of $P_{\rm F}$ with respect to $\overline{n}_{\rm j}$ in Figure~\ref{fig:pf_nj}. At very large scales $k\lesssim 0.007h\,{\rm Mpc}^{-1}$, the scale dependence of the $\overline n_{\rm j}$ is very different from the $b_{\rm j}$ effect; however this scale is outside the scale cut because of the radial width of the redshift slice.
Cross-correlating the \lya\ flux field with the galaxy field can help break this degeneracy.
 
\subsection{Constraints on Parameters}

We list the parameter fiducial values, priors and forecasted 1$\sigma$ uncertainties at $z = 2.4$, with fiducial overlapping area $\Omega_{\rm cross}=415\,\rm deg^2$ in Table~\ref{tb:params}. We group the parameters by whether they are totally determined by prior knowledge. As shown in the table, the constraints on $\{p_{\rm clump},\kappa_{\rm HI},  b_{\rm HI,u}, \beta_{\rm r}, b_{\rm clump}\}$ are mostly determined by their priors and can be hardly improved by this observation strategy. On the other hand, the survey indeed puts constraints on the two ionizing source parameters $b_{\rm j}$, $1/\overline{n}_{\rm j}$, two \lya\ flux bias  $b_{\rm F\delta}$, $b_{\eta}$ and the galaxy bias $b_{\rm g}$.

To explore possible survey strategy optimization that could be implemented by DESI or Roman, we plot the forecast error bars on the parameters with respect to overlapping survey area $\Omega_{\rm cross}$ and the two different Roman galaxy survey exposure durations in Figure~\ref{fig:param_opt}. For the fiducial galaxy survey scenario, we find from no overlap $(\Omega_{\rm cross}=0 \, \rm deg^2)$ to maximum overlapping area $(\Omega_{\rm cross}=2227 \, \rm deg^2)$ where the Roman footprint is fully contained within DESI, the detection sensitivity improvement for $b_{\rm j}$ is 37.2\% from 0.653 to 0.410, for $1/\overline{n}_{\rm j}$ it is 76.7\% from $3.814\times10^4h^{-3}\,\rm Mpc^3$ to $0.888\times10^4h^{-3}\,\rm Mpc^3$. Due to the reduction of cosmic variance by cross-correlation, and the breaking of partial degeneracies\footnote{The correlation coefficients Corr$(b_{\rm F\delta},1/\overline n_{\rm j})$ and Corr$(b_{\eta},1/\overline n_{\rm j})$ are +0.82 and $-0.85$, respectively, for the disjoint case; for the maximal overlap case, these are reduced to +0.22 and $-0.22$.} with the source shot noise $1/\overline n_{\rm j}$, the constraints on the biasing parameters $b_{\rm F\delta}$ and $b_{\eta}$ are also mildly improved by 45.0\% (0.00151 to 0.00083) and 48.9\% (0.00230 to 0.00117) respectively.

The breaking of the degeneracy between $b_{\rm j}$ and $1/\overline n_{\rm j}$ is shown in Figure~\ref{fig:contour}. The ellipses show the successive decrease in the forecast 68\% error ellipse as the overlap region between the two surveys is increased. The ellipse is of course a projection of a higher-dimensional ellipsoid in parameter space; as one goes beyond $\sim 1000$ deg$^2$ of overlap, other directions not constrained by the cross-correlation dominate the error region in the $(b_{\rm j},1/\overline n_{\rm j})$ plane, and there is only slight further improvement in the marginalized parameters. The dotted line in Fig.~\ref{fig:contour} shows the boundary of the unphysical region ($1/\overline{n}_{\rm j}<0$). In a Markov Chain Monte Carlo parameter analyses, one could add a step function prior on $1/\overline{n}_{\rm j}$ to exclude the unphysical values.\footnote{This could also be done in post-processing, since addition of a step function prior is equivalent to eliminating the unphysical region and re-normalizing the posterior probability to integrate to unity.}
However, the use of asymmetric priors could at the same time induce biased inference for other parameters, e.g., it is more likely to have larger inferred value for $b_{\rm j}$ than smaller ones because of the orientation of the contour in Figure. \ref{fig:contour}.
To avoid such issues in our forecast, we prefer to show the ``full'' measurement error before any such prior on $1/\bar n_j$ is applied.

\subsection{Impact of scale cuts}\label{subsec:convergence}

We test the impact of scale cuts on our results by doubling $k_{\rm max}$ to $0.4h\,{\rm Mpc}^{-1}$, or halving $k_{\parallel,\rm min}$ to $k_{\parallel,\rm min}=0.007485h\,\rm Mpc^{-1}$. We present the resulting changes in sensitivity in Table~\ref{tb:params}.

We find the constraining power improves when extending $k_{\rm max}$. Even though the UVB parameters will be highly degenerate at the high-$k$ regime because UVB fluctuation is a large-scale effect. The UVB parameters constraints changes a bit (7.2\% for $b_{\rm j}$ and 13.4\% for $1/\overline{n}_{\rm j}$). This is because the number of modes increases proportional to $k^3$ and thus there is still non-negligible information contributed by higher k modes to the Fisher matrix. Other parameters get better constraints (25.2\%, 31.5\%, 34.7\% for $b_{\rm F\delta}$, $b_{\eta}$, $b_{\rm g}$ respectively) simply because they have information from smaller-scale modes. 

 With access to larger-scale modes while $k_{\parallel,\rm min}$ is reduced in our convergence test, the degeneracy of UVB parameters is broken greatly as $b_{\rm j}$ constraint improved 7.0\% and $1/\overline{n}_{\rm j}$ improved 44.1\%. We note that for shorter quasar lifetimes, the shot noise at small $k$ is reduced, so in that case the extension to smaller $k_{\parallel,\rm min}$ may lead to a smaller improvement. The improvement for $b_{\rm F\delta}$ (21.3\%), $b_{\eta}$ (14.4\%), $b_{\rm g}$ (2.9\%) is not as obvious as with the information from small-scale modes.

In this paper, we show our final results in the scenario $k_{\rm max}=0.2h\, {\rm Mpc}^{-1}$, which will be safely outside the domain of non-linear effects.

\section{Discussion}
\label{sec:conclusion}

\begin{figure}
    \centering
    \includegraphics[width=3in]{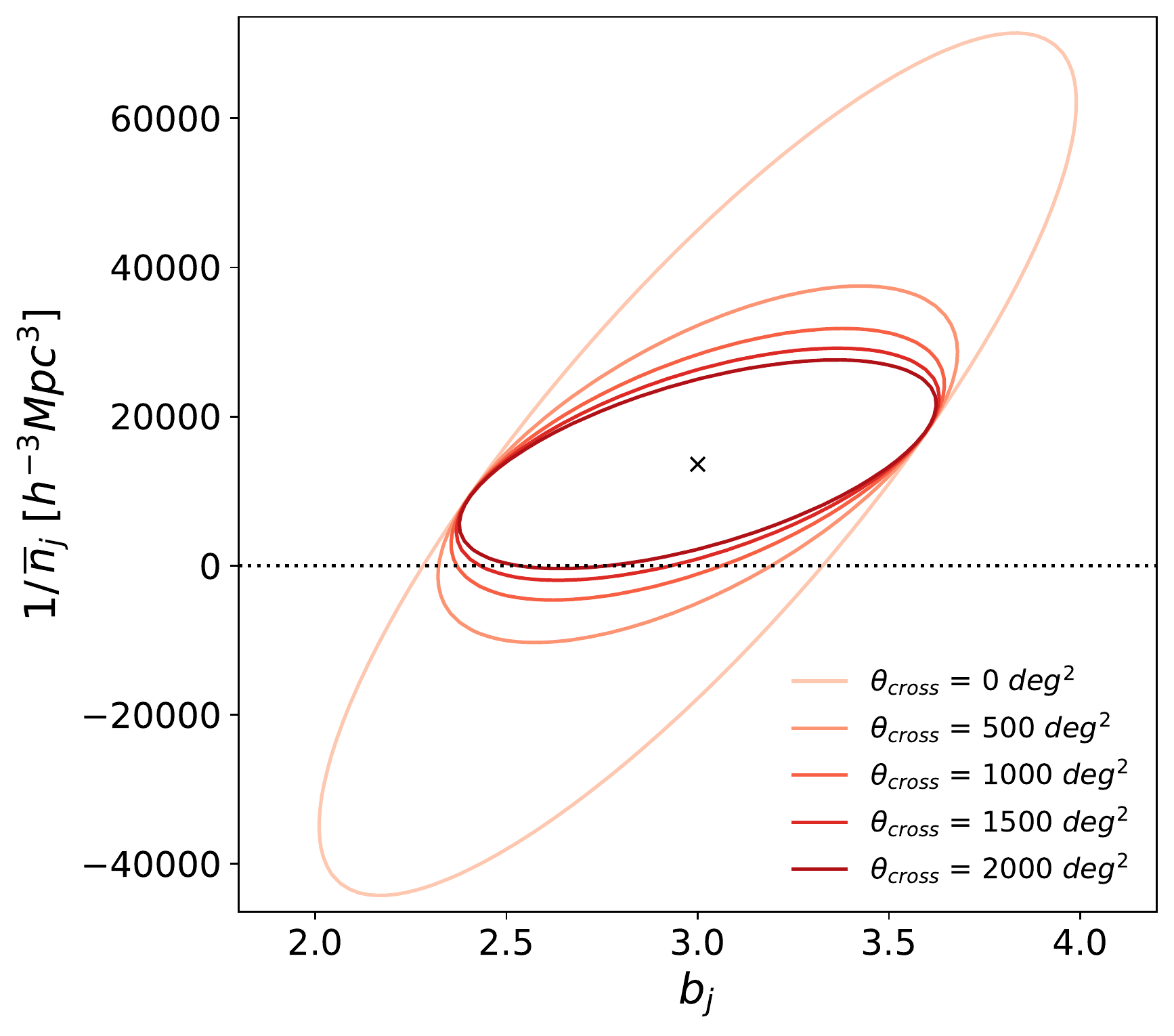}
    \caption{Projected 68\% confidence ($\Delta\chi^2=2.3$ for 2 degrees of freedom) contours of $b_{\rm j}$ and $1/\,\overline{n}_{\rm j}$. By increasing the overlapping area of DESI \lya\ forest survey and Roman galaxy survey, the degeneracy between these two parameters could be reduced. The dotted line at $1/\bar n_j$ shows the boundary of the physical region: the portion of the model space below it is unphysical.}
    \label{fig:contour}
\end{figure}

This work proposes to measure large-scale fluctuation of the ionizing background due to ionizing source distribution with a cross-correlation between \lya\ forest and a galaxy survey. We make a first estimate of the cross-correlation constraints on the ionizng sources bias $b_{\rm j}$ and mean number density $1/\overline{n}_{\rm j}$. We use the model in \cite{2014PhRvD..89h3010P} to parameterize the ionizing source distribution in the calculation of \lya\ flux power spectrum. The ionizing sources impact the \lya\ flux spectrum at large scales in two ways. First, at scales larger than the mean free path of ionizing photons, the clustering of ionizing sources will compete with the clustering of matter, enhance flux transmission and suppress the power spectrum because the absolute value of \HI\ bias is reduced. This effect is parameterized by $b_{\rm j}$ and illustrated in Figure~\ref{fig:pf_bj}. Second, the shot noise of ionizing source number density, parameterized by $\overline{n}_{\rm j}$, will give rise to very large fluctuation and donminate the power spectrum at extreme large scale (at the far left of Figure. \ref{fig:pf_nj}). These two parameters are somewhat degenerate because they both have their main effect on large scales, although the degeneracy is not exact because the $k$-dependence of the shot noise is steeper. Cross-correlating \lya\ flux with galaxy survey could help break this degeneracy since it could remove the shot noise term, as shown in Figure. \ref{fig:contour}.

\begin{figure*}
    \centering
    \includegraphics[width=6in]{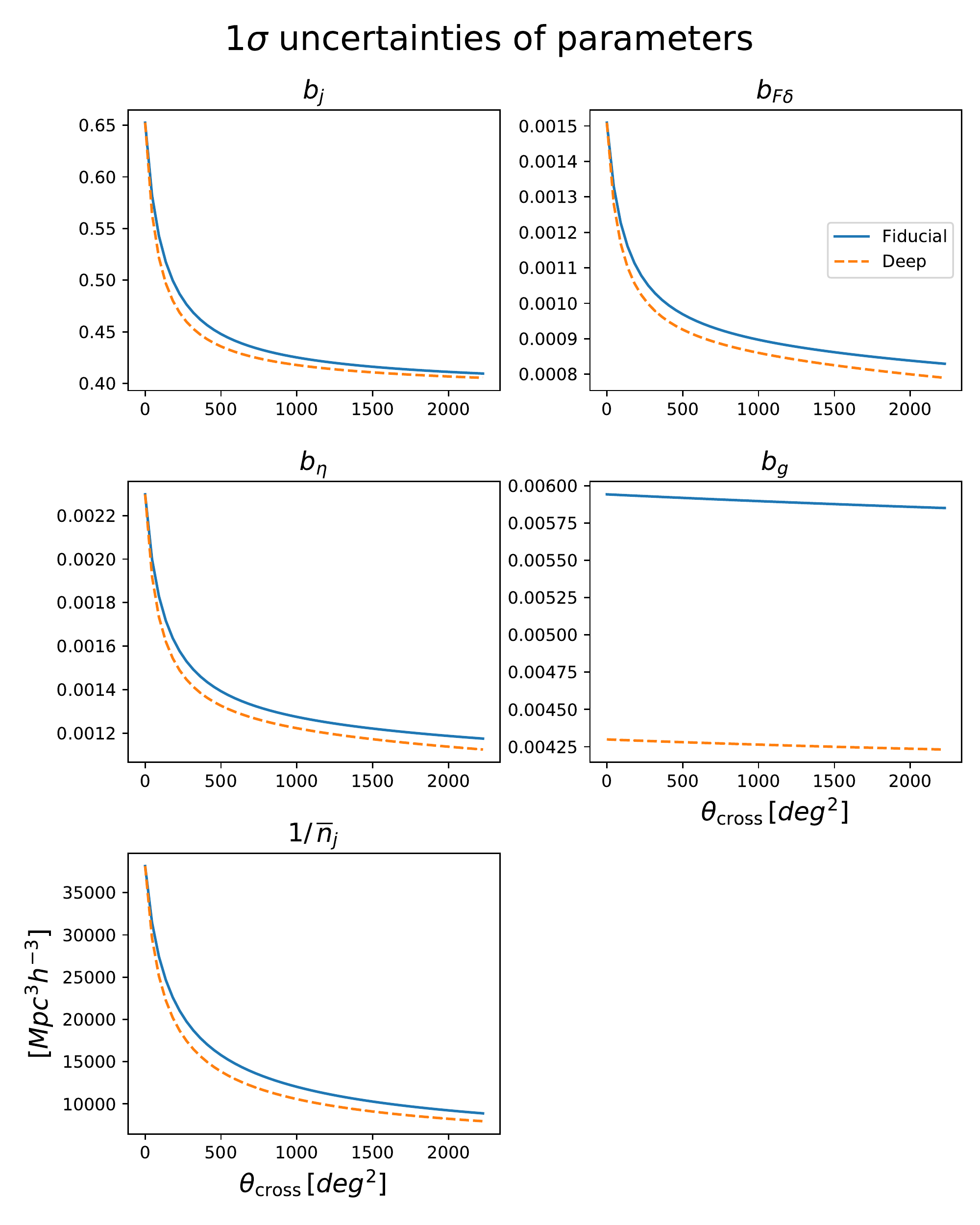}
    \caption{The 1$\sigma$ parameter uncertainties as a function of $\theta_{\rm cross}$, the overlap area of the DESI \lya\ forest survey and Roman galaxy survey.}
    \label{fig:param_opt}
\end{figure*}
We list the forecast constraints on our model parameters by the surveys in Table~\ref{tb:params}. We find two ionizing source related parameters $b_{\rm j}$ and $1/\overline{n}_{\rm j}$ could be constrained by \lya\ forest survey. Comparing the zero overlapping and maximum overlapping area between DESI \lya\ forest survey and Roman HLS, we find cross correlating with galaxy survey could improve the constraints, 37.2\% for $b_{\rm j}$ and 76.7\% for $1/\overline{n}_{\rm j}$ respectively. A deeper Roman galaxy survey could also improve the constraints mildly (38\% for $b_{\rm j}$ and 77 \% for $1/\overline{n}_{\rm j}$), although the overlap area with DESI is more important.

The model we investigate in this paper contains several simplifications, while future work could extend it to involve more complicated phenomena present in IGM. The ionizing source parameter $b_{\rm j}$ and $1/\overline{n}_{\rm j}$ are both effective numbers, mixing the contributions from all source populations. In future work, assigning unique parameters to distinct source population could help identify the fraction the contribution from each of them. There are also other factors that give rise to large scale fluctuation of the IGM opacity, such as temperature fluctuation as relics of \ion Hi/\ion{He}i reionization \citep[e.g.][]{2019MNRAS.487.1047M} and \ion{He}{ii} reionization \citep{2009ApJ...694..842M, 2013MNRAS.435.3169C, 2015MNRAS.447.2503G}. We leave the incorporation of these effects on large scale feature of IGM to future work. A full combined model of the UVB fluctuations may have to go beyond analytic approximations even at the largest scales and require a high dynamic range hydrodynamic simulation (e.g., as done in \citealt{2022arXiv220713098P} for patchy reionization) or hybrid scheme implemented on an $N$-body simulation (e.g., an extension of \citealt{2022MNRAS.514.3222P}), which could be an exploration direction for further work. In any case, with the inclusion of these additional complications, it is likely that one would try to consider more correlations (e.g., including the Lyman-$\alpha$ forest, Lyman-break and IR-selected galaxies, {\em and} quasars) to see if there is a possibility of breaking the added degeneracies. The addition of other statistics such as the 3-point correlation function of the \lya\ forest \citep{2019MNRAS.487.5346T} might also be useful, especially in the squeezed limit where one might expect a different configuration dependence if the low-$k$ mode is being affected by UVB fluctuations from source shot noise (the UVB is a scalar on small scales) versus a deviation correlated with the density field (whose non-linear couplings to small scales have all of the contributions in the second-order perturbation theory kernel, e.g., \citealt{1984ApJ...279..499F}).

On the more observational/data processing side, our model contained a rather simplistic approach to the cuts at the largest scales. In agreement with the previous study by \citet{2014ApJ...792L..34P}, we find that the treatment of these very large scales and what modes are projected out matters. A direct comparison is difficult since \citet{2014ApJ...792L..34P} worked in real space and we work in Fourier space. With $\kappa_{\rm HI}$ fixed and no marginalization over observational broadband parameters, they found $\sigma(b_{\rm j})\sim 0.12$ with Lyman-$\alpha$ alone (or $\sim 0.15$ scaled to our $\Delta z$ bin), a factor of $\sim 4$ smaller than we find here. However, we also note that our cut is in $k_{\parallel}$, so the large-scale modes that we keep have large $\mu$ and thus have the largest power from density+velocity fluctuations; we have marginalized over some additional parameters; there are some subtle differences in the way the \citet{2014PhRvD..89h3010P} model for intergalactic \ion Hi fluctuations was mapped onto the Lyman-$\alpha$ forest transmission (although both models include $b_{F\delta}$ and $b_\eta$); and the construction of the covariance matrix in the largest-scale bins is based on different assumptions. Future work should take into account the precise way in which the Lyman-$\alpha$ data are processed in order to predict which modes at the very largest scales are actually recovered. Our finding that these cuts matter further motivates work to model the quasar continuum and other large-scale observational systematics to maximize sensitivity to UVB physics.

For the future, other galaxy redshift surveys have been proposed that could explore this redshift range, such as the MaunaKea Spectroscopic Explorer \citep{2019arXiv190404907T} and MegaMapper \citep{2019BAAS...51g.229S}. The basic approach of cross-correlating the Lyman-$\alpha$ forest with galaxy samples would be applicable to these surveys as well, although the model would have to be expanded to take into account the unique features of the other target classes selected by the Lyman break. For example, on the very large scales of interest here, where $2\pi/k_\parallel\sim 400h^{-1}\,$Mpc is of order the separation of Lyman-$\alpha$ and Lyman-$\beta$, the shape and central rest frame wavelength of the Lyman break depends on the IGM density. Therefore one would have to include in the model a line-of-sight biasing term corresponding to how the change in transmission maps onto the color cuts used. We do not anticipate any fundamental difficulties in adding this to the model, but it is beyond the scope of this work.

Our results motivate survey strategies more dedicated to probe the UVB fluctuations.  We emphasize that the improvement shown in our table is just from choosing the footprint of the two surveys, neither of which has the UVB as its primary science case. There are of course practical limitations: DESI is based in the Northern Hemisphere, and especially at the bluer wavelengths used for the Lyman-$\alpha$ forest there is a loss of throughput as one points closer to the horizon. Roman is space-based, but the Reference Survey design \citep{2022ApJ...928....1W} was driven by the need to overlap with Southern Hemisphere telescopes such as the Vera Rubin Observatory\footnote{URL: https://www.lsst.org}, and the sky brightness (dominated by zodiacal light) increases as one moves closer to the Ecliptic \citep{1998A&AS..127....1L}. The survey choices may evolve in the future as the actual survey plan for Roman is defined, and as follow-on programs are considered for DESI. The large-scale ultraviolet background fluctuations are one example illustrating how the combined footprints of surveys should be considered in assessing the overall science reach.

\section*{Acknowledgements}

We thank Naim G\"oksel Kara{\c c}ayl{\i}, Andrei Cuceu, Paulo Montero-Camacho, and Paul Martini for useful discussion and comments in the preparation of this draft. We thank the anonymous referee for comments that improved the paper, especially regarding the finite source lifetime effects. During the preparation of this work, the authors were supported by NASA award 15-WFIRST15-0008, Simons Foundation award 60052667, and the David \& Lucile Packard Foundation.

\section*{DATA AVAILABILITY}
The data underlying this article will be shared on reasonable request to the corresponding author.

\bibliographystyle{mnras.bst}
\bibliography{main.bib}

\appendix

\section{Projecting a polynomial fit from a random field}
\label{app:fd}

In this appendix, we present a proof of Eq.~(\ref{eq:fD}), the reduction factor $f_D$ in variance of a 1-dimensional random field when the least squares fit polynomial is projected out.

First, let us consider a real random field $\psi(x)$ (not necessarily Gaussian), where $x$ is a position. In our application, $\psi$ is the transmitted flux perturbation $\delta_F$, but the result is more general. The field has a 1D power spectrum $P_\psi(k)$, defined in the usual way by
\begin{equation}
\langle \tilde\psi^\ast(k) \psi(k') \rangle = 2\pi \delta(k-k') P_\psi(k),
\label{eq:PA}
\end{equation}
and a variance
\begin{equation}
{\rm Var}[\psi(x)]
= \frac1\pi \int_0^\infty P_\psi(k)\,{\rm d}k
\label{eq:varpsi}
\end{equation}
that does not depend on position $x$.

Now we take a skewer of length $L$, which without loss of generality can be taken to extend over $-\frac12L<x<\frac12L$. We will find it convenient to expand $\psi$ in the basis of Legendre polynomials\footnote{Since $P$ and $L$ have other uses in this paper, we use ${\mathbb P}$ to denote the Legendre polynomials.} ${\mathbb P}_\ell(2x/L)$ since they are orthogonal over the skewer:
\begin{equation}
\int_{-L/2}^{L/2} {\mathbb P}_\ell\left(\frac{2x}L \right) {\mathbb P}_{\ell'}\left(\frac{2x}L \right) \,{\rm d}x = \frac{L}{2\ell+1}\delta_{\ell\ell'}.
\label{eq:P-ortho}
\end{equation}
Within the skewer, the function $\psi$ can be expressed as
\begin{equation}
\psi(x) = \sum_{\ell=0}^\infty a_\ell\, {\mathbb P}_\ell\left(\frac{2x}L \right), ~~~
-\frac12L<x<\frac12L,
\end{equation}
with
\begin{equation}
a_\ell = \frac{2\ell+1}{L} \int_{-L/2}^{L/2} \psi(x) {\mathbb P}_\ell\left(\frac{2x}L \right)\,{\rm d}x.
\end{equation}
We now want to find the polynomial $\psi_{\rm fit}$ of order $N$ that fits $\psi$ with the smallest mean square residual. We write the polynomial as $\psi_{\rm fit}(x) = \sum_{\ell=0}^N b_\ell {\mathbb P}_\ell(2x/L)$, where the $N+1$ coefficients $b_0 ... b_N$ float. The mean square residual is
\begin{eqnarray}
{\rm MSR} \!\!\!\! &=& \!\!\frac1L \int_{-L/2}^{L/2} [\psi(x) - \psi_{\rm fit}(x)]^2 \,{\rm d}x
\nonumber \\
&=& \!\!\frac1L \int_{-L/2}^{L/2} \left[
\sum_{\ell=0}^\infty a_\ell\, {\mathbb P}_\ell\left(\frac{2x}L \right) - \sum_{\ell=0}^N b_\ell\, {\mathbb P}_\ell\left(\frac{2x}L \right)\right]^2\,{\rm d}x
\nonumber \\
&=&\!\! \sum_{\ell=0}^N \frac{(a_\ell-b_\ell)^2 }{2\ell+1} + \sum_{\ell=N+1}^\infty \frac{a_\ell^2}{2\ell+1},
\end{eqnarray}
where we used the orthogonality relation (Eq.~\ref{eq:P-ortho}) in the last step. The minimum is achieved when $b_\ell=a_\ell$ for $\ell = 0,1,...N$, since this sets the first term to its minimum possible value of zero. Then the mean square residual is simply the second term. Its expectation value is
\begin{eqnarray}
\langle {\rm MSR} \rangle \!\!\!\! &=& \!\!
\sum_{\ell=N+1}^\infty \frac{\langle a_\ell^2\rangle}{2\ell+1}
\nonumber \\
&=& \!\! \sum_{\ell=N+1}^\infty  \frac{2\ell+1}{L^2}
\left\langle\left| \int_{-L/2}^{L/2} \psi(x) {\mathbb P}_\ell\left( \frac{2x}L \right)\,{\rm d}x
\right|^2 \right\rangle
\nonumber \\
&=& \!\! \sum_{\ell=N+1}^\infty  \frac{2\ell+1}{L^2}
\Biggl\langle\Biggl| \int_{-\infty}^\infty \frac{{\rm d}k}{2\pi}\, \tilde\psi(k) 
\nonumber \\ && ~~~~\times
\int_{-L/2}^{L/2} {\rm d}x \, {\rm e}^{{\rm i}kx} {\mathbb P}_\ell\left( \frac{2x}L \right)
\Biggr|^2\Biggr\rangle
\nonumber \\
&=& \!\! \sum_{\ell=N+1}^\infty  \frac{2\ell+1}{L^2}
\Biggl\langle\Biggl| \int_{-\infty}^\infty \frac{{\rm d}k}{2\pi}\, \tilde\psi(k) 
\, {\rm i}^\ell L\, j_\ell\left(\frac{kL}2\right)
\Biggr|^2\Biggr\rangle
\nonumber \\
&=& \!\! \sum_{\ell=N+1}^\infty  (2\ell+1)
\int_{0}^\infty \frac{{\rm d}k}{\pi} \, P_\psi(k) \left[j_\ell\left(\frac{kL}2\right)\right]^2
,
\label{eq:MSR}
\end{eqnarray}
where in the second equality we used the fact that $a_\ell$ is real; in the fourth equality we used the integral relating the Legendre polynomials to the spherical Bessel functions, Eq.~(10.1.14) of \citet{1972hmfw.book.....A}; and in the final equality we used the expectation value (Eq.~\ref{eq:PA}), used the $\delta$-function to collapse one of the two copies of the $k$-integral, and used the even nature of the integrand to write the integral over only positive $k$. We may then write the ratio of the expected mean square residual to the variance of the original field:
\begin{equation}
\frac{\langle{\rm MSR}\rangle}{{\rm Var}[\psi(x)]}
= \frac{\int_0^\infty P_\psi(k) f_D(k) \,{\rm d}k/\pi}{\int_0^\infty P_\psi(k) \,{\rm d}k/\pi},
\end{equation}
where the $k$-dependent reduction factor $f_D(k)$ is
\begin{equation}
f_D(k) = \sum_{\ell=N+1}^\infty  (2\ell+1)
\left[j_\ell\left(\frac{kL}2\right)\right]^2.
\label{eq:red}
\end{equation}
The sum rule, Eq.~(10.1.50) of \citet{1972hmfw.book.....A}, allows us to write this instead as
\begin{equation}
f_D(k) = 1 - \sum_{\ell=0}^N  (2\ell+1)
\left[j_\ell\left(\frac{kL}2\right)\right]^2,
\end{equation}
which completes the proof of Eq.~(\ref{eq:fD}). Note that from Eq.~(\ref{eq:red}), $f_D(k) \rightarrow 0$ as $N\rightarrow\infty$.

Another useful limiting form is that for small $kL$, the Taylor series for $j_\ell$ (Eq.~10.1.2 of \citealt{1972hmfw.book.....A}) gives $f_D\propto (kL)^{2(N+1)}$. In particular, for $N=1$ (linear fit) we have $f_D(k) \rightarrow \frac1{720}(kL)^4$.

\section{2-dimensional clump estimation of fractional opacity}
\label{appendix:clump}
\begin{figure}
    \centering
    \includegraphics[width=1\columnwidth]{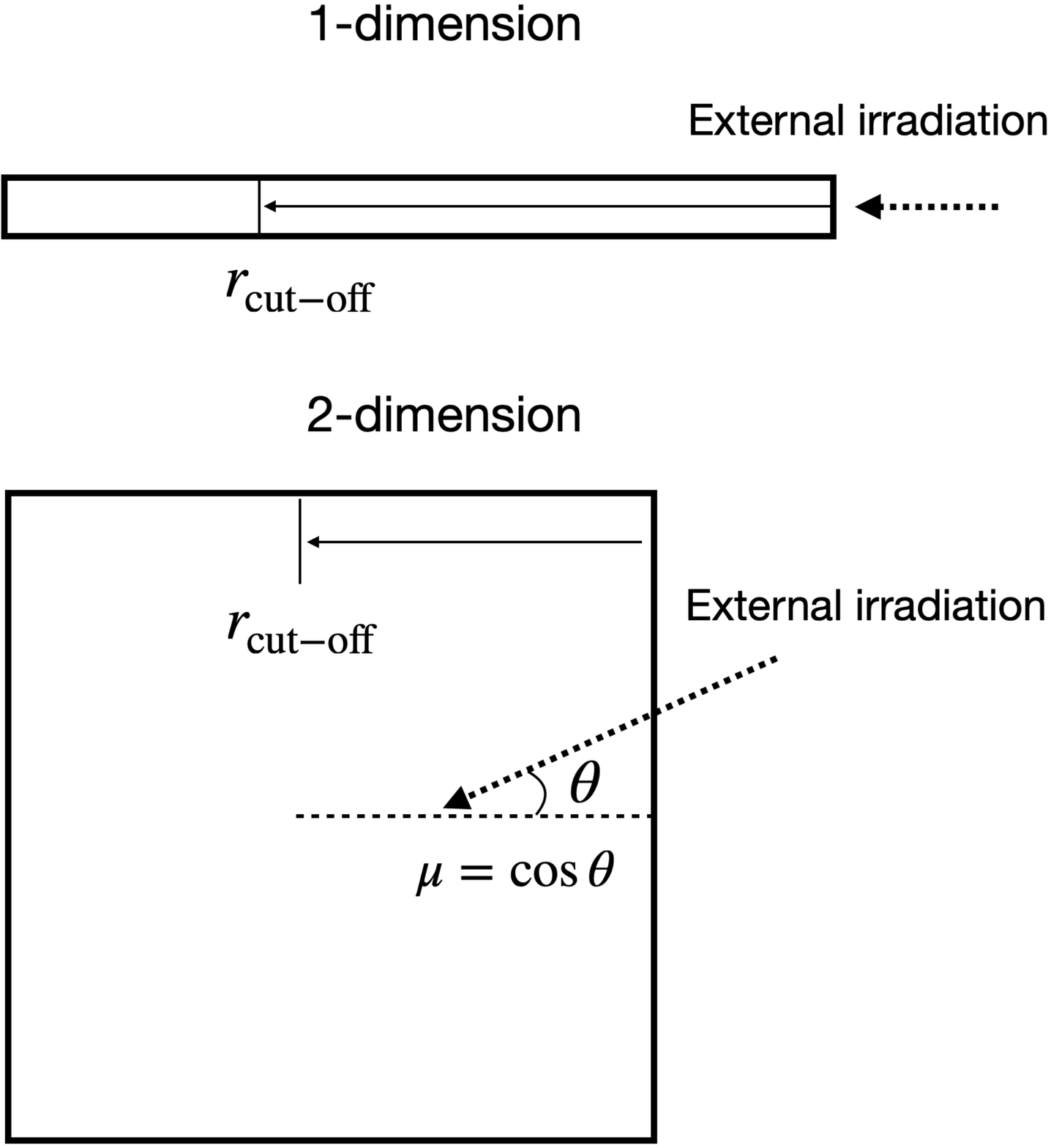}
    \caption{Illustrations for the 1D and 2D clump models we use to estimate the fraction of opacity by clumping regions and IGM \HI\ .}
    \label{fig:clump}
\end{figure}

We illustrate our 1-dimensional and 2-dimensinal clump models (used in the estimation of $p_{\rm clump}$ in the main text) in Figure~\ref{fig:clump}. The photionization rate $\Gamma_{pi}$ decays exponentially with the total hydrogen column density $N$. When irradiated by an external flux, the mean $\Gamma_{pi}$ can be approximately written as
\begin{equation}
    \Gamma_{pi} \propto\Gamma_0\left\langle e^{-\overline{\sigma}_{\rm HI}N(\textbf{r})} \right\rangle,
\end{equation}
where $\Gamma_0$ is the phoionization rate in the exterior. 

The 1-dimensional model, most similar to the method of \citet{2014ApJ...792L..34P}, considers radiation only propagating along a single axis, which is also used to measure the column density.
Converting the depth $r$ inside of the clump to the fraction of cut-off column density $N_{\rm cut-off}$, we get an estimation for the 1-dimensional clump model as follows
\begin{eqnarray}
  \left\langle e^{-\sigma_{\rm HI}N(r)} \right\rangle
  &=& \int_0^1 dx\, e^{-\sigma_{\rm HI}N_{\rm cut-off}x(r)}
  \nonumber \\
  &=& \frac{1-e^{-\overline{\sigma}_{\rm HI}N_{\rm cut-off}}}{\overline{\sigma}_{\rm HI}N_{\rm cut-off}}
  \nonumber \\
  &\approx&\frac{1}{\overline{\sigma}_{\rm HI}N_{\rm cut-off}},
\end{eqnarray}
where the coordinate $x$ ($0<x<1$) is the fraction of the path through the cloud ($x=0$ at the exposed surface).

For the 2-dimensional model, there is an additional dependence on the external irradiation angle $\theta$, which is taken to be isotropic (uniformly distributed in $\mu=\cos\theta$):
\begin{eqnarray}
  \left\langle e^{-\overline{\sigma}_{\rm HI}N(r,\mu)} \right\rangle 
  &=& \int_0^1dx\, \int_0^1 d\mu\,e^{-\sigma_{\rm HI}N_{\rm cut-off}x(r)/\mu}\nonumber \\
  &=& \int_0^1 d\mu\,\frac{1-e^{-\overline{\sigma}_{\rm HI}N_{\rm cut-off}/\mu}}{\overline{\sigma}_{\rm HI}N_{\rm cut-off}/\mu} \nonumber \\
  &\approx& \frac{1}{2\overline{\sigma}_{\rm HI}N_{\rm cut-off}}.
\end{eqnarray}
This is different by a factor of 2.

We emphasize that both of these models refer to highly oversimplified geometries; their main purpose is to show how much variation in the relation between $\Gamma_{pi}$ and $N_{\rm cut-off}$ can be obtained by order-unity factors in the geometry of the cloud, and hence what is a reasonable prior uncertainty in $p_{\rm clump}$.

\section{Source lifetimes and the shot noise term}
\label{app:variability}

Finite source lifetimes can reduce the quasar shot noise term in ionizing background models: if the scale of the perturbation $1/k$ is large compared to the distance a photon can travel during the quasar lifetime $ct_{\rm Q}/a$ (where $t_{\rm Q}$ is the quasar lifetime), then there is an effective increase in the number of individual sources that contribute to the ionizing background. \citet{2014MNRAS.442..187G} modeled this in correlation function space by multiplying the shot noise term by a scale dependent factor. \citet{2019MNRAS.482.4777M} performed a sophisticated Fourier-space calculation of the time-dependent effects. The purpose of this appendix is to provide a simple analytic estimate of the finite lifetime effects that is usable in the range of scales most of interest here ($k\sim$ few$\times 10^{-2}h$ Mpc$^{-1}$). We would like to include the proper geometric factors that were alluded to in \citet{2014MNRAS.442..187G}, while having an expression that is simple to understand.

The basic question we seek to answer here is the following: suppose there is a set of sources with some mean number density $1/\bar n$ at any given time. These sources emit radiation into the IGM that is absorbed (or redshifted) with some mean free path $\kappa_{\rm tot}^{-1}$. Suppose that there is a probability $p(\Delta t)$ that a source that is ``on'' at some time $t$ is still on at time $t+\Delta t$. By what factor $f_{\rm FL}(k,\mu)$ is the power spectrum of radiation intensity flucutations relative to a steady-state distribution of sources? We address this question at scales small compared to the Hubble scale, $k\gg aH$, so that we can have time-dependent perturbations in the radiation intensity on a time-independent background (in the sense described in \citealt{2019MNRAS.482.4777M}). We allow for a $\mu$-dependence here because we compute the power spectrum on the light cone, i.e., allowing for the variation in look-back time as seen by the observer.

\begin{figure}
\includegraphics[width=3.25in]{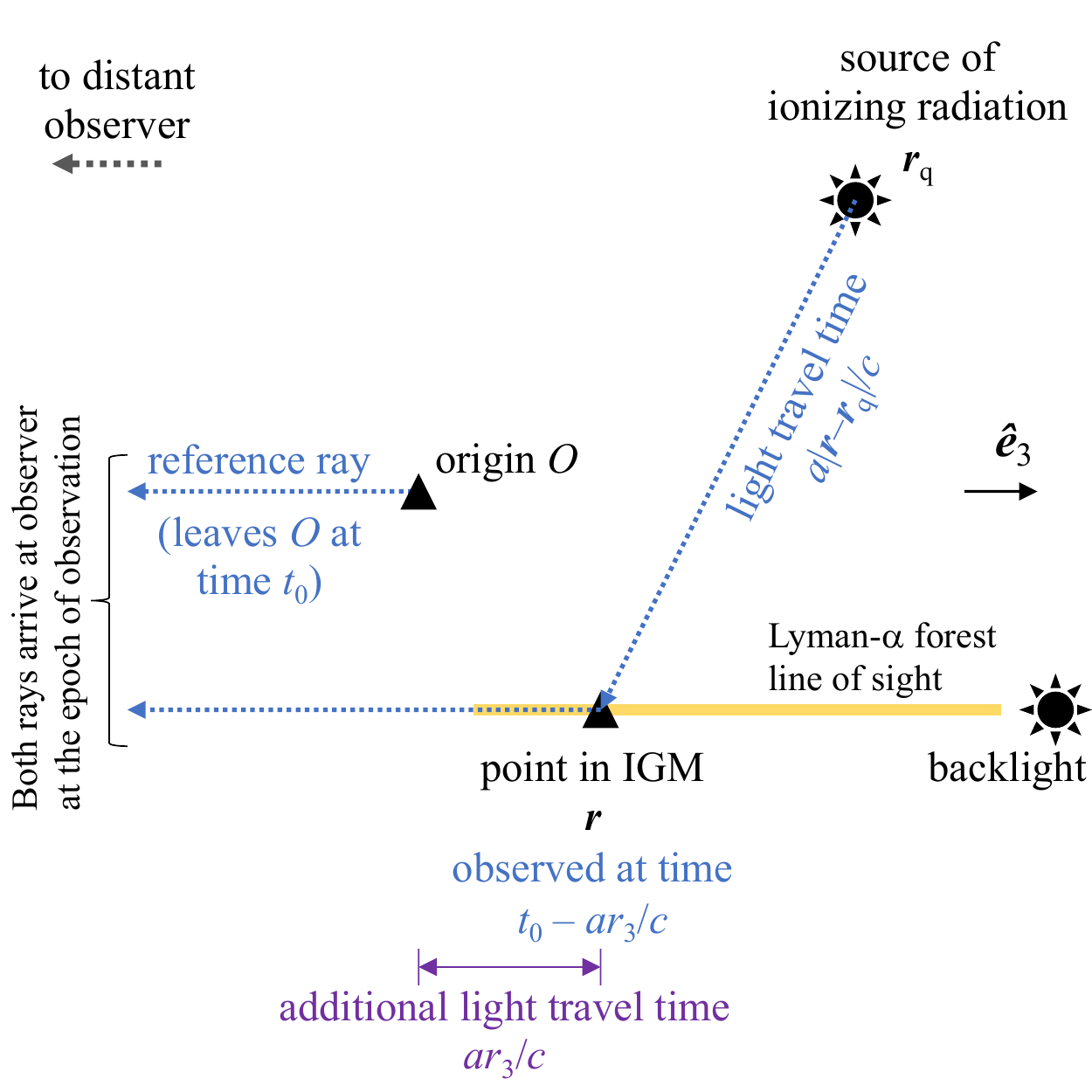}
\caption{\label{fig:tgeom}The geometry for the time-dependent shot noise calculation in Appendix~\ref{app:variability}, showing the symbols in Eq.~(\ref{eq:Jrt}). We work in the distant-observer approximation, with the observer looking to the right ($\hat{\bmath e}_3$ direction). We define an arbitrary location $O$ to be the origin; the observer sees $O$ as it was at some time $t_0$. A given point ${\bmath r}$ in the Lyman-$\alpha$ forest is seen as it was at time $t_0-ar_3/c$ due to the additional light travel time noted along the bottom path (note the factor of $a$ since we work in comoving position but physical time; and that $r_3$ is the component of ${\bmath r}$ in the line of sight direction). This point ${\bmath r}$ is illuminated by a source at position ${\bmath r}_q$, separated by a light-travel time $a|{\bmath r}-{\bmath r}_q|/c$. Overall, the time radiation must have left the source ${\bmath r}_q$ to contribute to the radiation intensity at ${\bmath r}$ at the relevant epoch for the distant observer is $t_0 - ar_3/c - a|{\bmath r}-{\bmath r}_q|/c$; therefore it is this time that appears in Eq.~(\ref{eq:Jrt}).}
\end{figure}

Let us suppose that the observer is looking along the $z$-axis (in direction $\hat{\bmath e}_3$). Suppose we consider a box of volume $V$, considering $N$ sources of luminosity $L$, where source $q$ are each on for some interval of time ${\cal B}_q$ (where $q=1...N$). Then the intensity of radiation at point ${\mathbf r}$ is
\begin{equation}
J({\bmath r},t_0) = \sum_q \frac{L {\rm e}^{-\kappa_{\rm tot} a|{\bmath r}-{\bmath r}_q|}}{4\pi a^2 |{\bmath r}-{\bmath r}_q|^2 }\; \chi_{{\cal B}_q}\left(t_0 - a\frac{r_3+|{\bmath r}-{\bmath r}_q|}c\right),
\label{eq:Jrt}
\end{equation}
where $t_0$ is the time at which light must leave the origin to reach the observer; the sum is over sources; $a|{\bmath r}-{\bmath r}_q|$ is the physical distance from the point of interest to the source; the inverse square law and exponential attenuation have been considered; $\chi_{{\cal B}_q}$ is the characteristic function that is 1 if the argument is in the interval ${\cal B}_q$ and 0 otherwise; and the argument includes the point-to-observer delay and the source-to-point delay (see Fig.~\ref{fig:tgeom}).

Equation~(\ref{eq:Jrt}) can be simplified by using the Fourier transform of the characteristic function:
\begin{equation}
\chi_{{\cal B}_q}(t) = \int_{-\infty}^\infty 
\tilde\chi_{{\cal B}_q}(\omega) {\rm e}^{-{\rm i}\omega t} \,\frac{{\rm d}\omega}{2\pi}.
\end{equation}
The characteristic function has the correlation function
\begin{equation}
\langle \chi_{{\cal B}_q}(t) \chi_{{\cal B}_q}(t+\Delta t) \rangle
= \frac{\bar n V}{N}p(\Delta t),
\end{equation}
and so its frequency-space power spectrum is
\begin{equation}
\langle \tilde\chi_{{\cal B}_q}^\ast(\omega) \tilde\chi_{{\cal B}_q}(\omega') \rangle
= 2\pi \frac{\bar n V}{N} \tilde p(\omega)\delta(\omega-\omega'),
\label{eq:fps}
\end{equation}
where $\tilde p$ denotes the Fourier transform of $p$. The steady source case would have $p(\Delta t)=1$ and $\tilde p(\omega) = 2\pi \delta(\omega)$.

We also take the 3-dimensional Fourier transform of $J$ at fixed $t_0$:
\begin{eqnarray}
\tilde J({\bmath k},t_0) \!\!\!\! &=& \!\!\!\!
\int_V J({\bmath r},t_0) \,{\rm e}^{-{\rm i}{\bmath k}\cdot{\bmath r}}\,{\rm d}^3{\bmath r}
\nonumber \\
&=& \!\!\!\!
\int_{-\infty}^\infty \sum_q \int_V \frac{L {\rm e}^{-\kappa_{\rm tot} as_q}}{4\pi a^2 s_q^2 }\; 
{\rm e}^{-{\rm i}\omega [t_0 - a(r_3 + s_q)/c]}
\tilde\chi_{{\cal B}_q}(\omega)
\nonumber \\ && \times
{\rm e}^{-{\rm i}{\bmath k}\cdot{\bmath r}}\,{\rm d}^3{\bmath r}
\,\frac{{\rm d}\omega}{2\pi}
\nonumber \\
&=& \!\!\!\! L
\int_{-\infty}^\infty \frac{{\rm d}\omega}{2\pi} \sum_q 
{\rm e}^{-{\rm i}{\bmath k}\cdot{\bmath r}_q}
{\rm e}^{-{\rm i}\omega (t_0 - ar_{q3}/c)}
\tilde\chi_{{\cal B}_q}(\omega)
\nonumber \\ && \times
\int_V \,{\rm d}^3{\bmath s}_q
\frac{ {\rm e}^{-(\kappa_{\rm tot}-{\rm i}\omega/c) as_q}}{4\pi a^2 s_q^2 } 
{\rm e}^{-{\rm i}({\bmath k} - a \omega \hat{\bf e}_3/c)\cdot{\bmath s}_q}
,
\label{eq:Jk}
\end{eqnarray}
where we define ${\bmath s}_q \equiv {\bmath r} - {\bmath r}_q$, and in the last line changed the integration variable to ${\bmath s}_q$ and split apart the terms in the exponential.

The power spectrum of $J$ is then obtained by the usual procedure of taking the square norm of Eq.~(\ref{eq:Jk}), averaging, and dividing by $V$. Since we are considering the case of shot noise, we consider only the contributions to the square of Eq.~(\ref{eq:Jk}) arising from the same source. This means we can take a single source $q$ at a random position and multiply by $N$. The expectation values involving $\tilde\chi_{{\cal B}_q}(\omega)$ are simplified using Eq.~(\ref{eq:fps}) -- this cancels the arbitrary factors of $N$ and $V$, and the $\delta$-function collapses one of the frequency integrals. The result is:
\begin{eqnarray}
P_{\tilde J}({\bmath k}) \!\!\!\! &=& \!\!\!\!
\bar n L^2
\int_{-\infty}^\infty \frac{{\rm d}\omega}{2\pi}\, \tilde p(\omega)
\nonumber \\ && \times
\Bigl|\int_V \,{\rm d}^3{\bmath s}_q
\frac{ {\rm e}^{-(\kappa_{\rm tot}-{\rm i}\omega/c) as_q}}{4\pi a^2 s_q^2 } 
{\rm e}^{-{\rm i}({\bmath k} - a \omega \hat{\bf e}_3/c)\cdot{\bmath s}_q}\Bigr|^2.
\nonumber \\ &&
\label{eq:PJk}
\end{eqnarray}

The integral in the absolute value (which we will call ${\cal I}$) can be simplified by the usual procedure of writing it in polar coordinates. Let us define the vector
\begin{equation}
{\bmath K} \equiv {\bmath k} - \frac{a\omega}c \hat{\bmath e}_3.
\end{equation}
The magnitude is
\begin{equation}
K = \sqrt{k^2 + \frac{a^2\omega^2}{c^2} - 2\frac{a\omega}{c}k\mu}.
\end{equation}
Then the volume integral in Eq.~(\ref{eq:PJk}) can be written in polar coordinates with the ``North Pole'' pointed toward ${\bmath K}$. There is an angle from the North Pole, $\bar\theta$ (with $\bar\mu=\cos\bar\theta$; we will use a bar to distinguish this from line-of-sight angle of ${\bmath k}$) and a longitude $\bar\phi$ with trivial dependence that brings out a factor of $2\pi$. There is also a radial integral over $s_q$. Then the integral is:
\begin{eqnarray}
{\cal I} &=& \frac12 \int_{-1}^1 {\rm d}\bar\mu\,
\int_0^\infty {\rm d}s_q\,
\frac{ {\rm e}^{-(\kappa_{\rm tot}-{\rm i}\omega/c - {\rm i}K\bar\mu/a) as_q}}{a^2 } 
\nonumber \\
&=&
\frac1{2a^3} \int_{-1}^1 \frac{{\rm d}\bar\mu}{
\kappa_{\rm tot}-{\rm i}\omega/c - {\rm i}K\bar\mu/a }
\nonumber \\
&=&
\frac{{\rm i}}{2a^2K} 
\ln\frac
{\kappa_{\rm tot}-{\rm i}\omega/c - {\rm i}K/a }
{\kappa_{\rm tot}-{\rm i}\omega/c + {\rm i}K/a }
\nonumber \\
&=&
\frac{1}{2a^2K} 
\Biggl[
\tan^{-1} \frac{K/a+\omega/c}{\kappa_{\rm tot}}
+ \tan^{-1} \frac{K/a-\omega/c}{\kappa_{\rm tot}}
\nonumber \\ &&
+
\frac12{\rm i} \ln \frac{\kappa_{\rm tot}^2 + (K/a+\omega/c)^2}
{\kappa_{\rm tot}^2 + (K/a-\omega/c)^2}
\Biggr]
.
\end{eqnarray}
In the last line, we have used the usual rule for the logarithm of a complex number in terms of its polar coordinates. We then arrive at a final expression for the power spectrum:
\begin{equation}
P_{\tilde J}(k,\mu) = \frac{\bar n L^2}{\kappa_{\rm tot}^2 a^6} \int_{-\infty}^\infty \frac{{\rm d}\omega}{2\pi}\, 
\tilde p(\omega)\,{\mathcal H}\left( \frac{K}{a\kappa_{\rm tot}}, \frac{\omega}{\kappa_{\rm tot}c} \right),
\label{eq:PH}
\end{equation}
where we have defined the ${\mathcal H}$-function
\begin{eqnarray}
{\mathcal H}(x,y) \!\!\!\! &\equiv& \!\!\!\!\left[ \frac{\tan^{-1}(x+y) + \tan^{-1}(x-y)}{2x} \right]^2
\nonumber \\ &&
+ \frac1{16x^2} \left[ \ln \frac{1+(x+y)^2}{1+(x-y)^2}
\right]^2.
\label{eq:Hfunc}
\end{eqnarray}
We note the limiting case
\begin{equation}
{\mathcal H}(x,0) = \left(\frac{\tan^{-1}x}x\right)^2 = 
[S(a\kappa_{\rm tot}x)]^2,
\end{equation}
so in the time-independent case ($\omega=0$) we recover the usual result.

Now in the time-independent case, $\tilde p(\omega) = 2\pi\delta(\omega)$, the power spectrum then collapses to $\bar n L^2 [S(k)]^2/(\kappa_{\rm tot}^2 a^6)$. Dividing Eq.~(\ref{eq:PH}) by this factor, we arrive at the shot noise suppression factor due to finite lifetimes:
\begin{equation}
f_{\rm FL}(k,\mu) = \frac1{[S(k)]^2} \int_{-\infty}^\infty \frac{{\rm d}\omega}{2\pi}\, 
\tilde p(\omega)\,{\mathcal H}\left( \frac{K}{a\kappa_{\rm tot}}, \frac{\omega}{\kappa_{\rm tot}c} \right).
\label{eq:ffl}
\end{equation}
In the case that the quasars are ``on'' for a duration of time $t_{\rm Q}$, then the conditional probability of being on at a later time is a triangle function:
\begin{equation}
p(\Delta t) = \left\{ \begin{array}{lll}
1-|\Delta t|/t_{\rm Q} & & |\Delta t|<t_{\rm Q} \\
0 & & |\Delta t|\ge t_{\rm Q}
\end{array}\right.,
\end{equation}
so
\begin{equation}
\tilde p(\omega) = t_{\rm Q}\left[\frac{\sin (\omega t_{\rm Q}/2)}{(\omega t_{\rm Q}/2)} \right]^2.
\end{equation}

\begin{figure}
    \centering
    \includegraphics[width=3.2in]{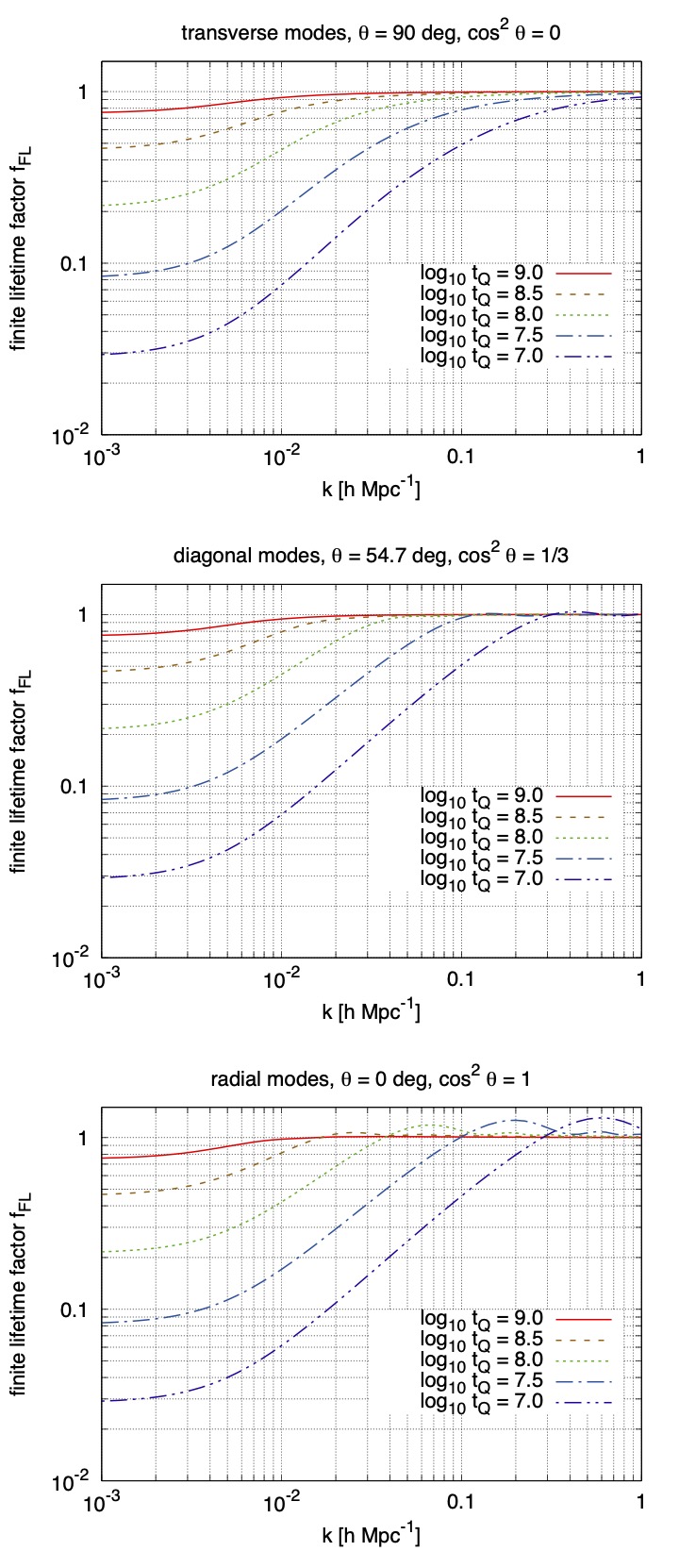}
    \caption{\label{fig:ffplot}The factor $f_{\rm FL}$ by which shot noise in the ionizing radiation field is suppressed by finite quasar lifetimes, as given by Eq.~(\ref{eq:ffl}). The top panel shows transverse modes ($\mu=0$); the middle panel shows diagonal modes at the root of the $P_2$ Legendre polynomial ($\mu=1/\sqrt3$); and the bottom panel shows radial modes ($\mu=1$). The five curves show quasar lifetimes ranging from $t_{\rm Q}=10^9$ yr (solid red, for illustration only since this is extremely long) through $10^7$ yr (dot-dot-dashed purple). }
\end{figure}

We cannot find an analytic solution to the integral in Eq.~(\ref{eq:ffl}). However, it is a one-dimensional integral of elementary functions and thus easily solvable numerically. Results are shown in Fig.~\ref{fig:ffplot}. We see that at large scales, there is a large suppression of the shot noise contribution to ionizing background fluctuations, especially for the shorter quasar lifetimes. For $k \ll a/(ct_{\rm Q})$, the suppression factor $f_{\rm FL}$ is proportional to $t_{\rm Q}$, consistent with the findings in \citet[\S2.3]{2019MNRAS.482.4777M}. At smaller scales (larger $k$), $f_{\rm FL}\rightarrow 1$ and the shot noise power is unaffected. There is a modest dependence on the orientation of the wave vector with respect to the line of sight. For radial modes (large $\mu$), a ``high-$k$ ringing'' is visible where the ionizing background power spectrum is actually enhanced for some modes. The peak of $f_{\rm FL}$ is at $k\approx 2\pi a/( ct_{\rm Q})$, where the radial comoving distance between photons that left the quasar when it turned on and reverberated off the IGM behind the quasar and the photons that left the quasar when it turned off and came directly toward the observer ($ct_{\rm Q}/2a$) is a half-wavelength. This feature moves to higher $k$ for shorter quasar lifetimes. The ringing makes sense given that the model assumed the same lifetime for all quasars, but would probably be smoothed out in a realistic scenario.

We see that the very large suppression factors $f_{\rm FL}$ are obtained at the lowest $k$, which we do not use in our analysis. At the largest scale we use, $k=0.015h\,${\rm Mpc}$^{-1}$ and $\mu=1$, the suppression factors can still be significant -- they are 0.085, 0.23, and 0.55 for quasar lifetimes of $10^7$, $10^{7.5}$, and $10^8$ yr respectively. So for long quasar lifetimes ($10^8$ yr), there is only a modest reduction in the shot noise factor, whereas for the short lifetimes the shot noise could be very dramatically reduced.

\end{document}